\documentclass[useAMS,usenatbib]{mn2e}
\usepackage{amsfonts,amssymb}
\usepackage{graphicx}

\usepackage{ulem}

\title{POPULATION SYNTHESIS FOR SYMBIOTIC STARS WITH WHITE DWARF ACCRETORS}
\author[ L\"{u},  Yungelson and   Han ]{Guoliang L\"{u}$^{1, 2, 3}$\thanks{E-mail:
ytlgl@yahoo.com.cn (LGL)}, L. Yungelson$^{4}$ and Z. Han$^{1}${}\\
$^{1}$National Astronomical Observatories / Yunnan Observatory,
the Chinese Academy of Sciences,
P.O.Box 110, Kunming, 650011, China \\
$^2$Graduate School of the Chinese Academy of Sciences, Beijing, China  \\
$^3$Department of Physics, Xinjiang University, Ulumuqi, 830046,
China \\
 $^{4}$Institute of Astronomy of the Russian Academy of
Sciences, 48 Pyatnitskaya Str., Moscow, Russia}
\begin{document}

\date{}

\pagerange{\pageref{firstpage}--\pageref{lastpage}} \pubyear{2006}

\maketitle

\label{firstpage}

\begin{abstract}
We have carried out a detailed study of symbiotic stars with white
dwarf accretors by means of a population synthesis code. We
estimate the total number of symbiotic stars with white dwarf
accretors in the Galaxy  as 1,200 -- 15,000. This range is
compatible with observational estimates. Two crucial physical
parameters that define the birthrate and number of symbiotic stars
are the efficiency of accretion by white dwarfs (which greatly
depends on the separation of components after common envelope
stage and stellar wind velocity) and the mass of the hydrogen
layer which the white dwarf can accumulate prior to the hydrogen
ignition. The theoretical estimate of the Galactic occurrence rate
of symbiotic novae ranges from about 1.3 to about 13.5 yr$^{-1}$,
out of which weak symbiotic novae comprise about 0.5 to 6.0
yr$^{-1}$, depending on the model assumptions. We simulate the
distributions of symbiotic stars over orbital periods, masses of
components, mass-loss rates of cool components, mass-accretion
rates of hot components and  luminosity of components. Agreement
with observations is reasonable.
\end{abstract}


\section{Introduction}

Symbiotic stars (SySs) are an inhomogeneous group of variable stars with
composite spectra. Their spectacular spectral and photometric variability is a
very important and interesting phenomenon. The spectra of SySs suggest that a
three-component system consists of a binary system in which an evolved giant
transfers matter to a much hotter compact companion by means of stellar wind
and an HII region \citep{l1,l2,l3}.

The cool component is a red giant (RG) which is a first giant branch (FGB) or
an asymptotic giant branch (AGB) star. In the majority of SySs the hot
component is, most probably, a white dwarf (WD), a subdwarf or an accreting
low-mass main-sequence (MS) star \citep{b50,b18,b32,b66,b16}. The variability
of SySs may be due to the thermonuclear runaways on the surface of an accreting
WD \citep{b50,b60,b16} or to the variations in the accretion rate onto the hot
component \citep{d86,l18,l13}. Recent reviews of the properties of SySs can be
found in \citet{b34} and \citet{b29}. The nova-like eruptions of SySs were
reviewed by \citet{b28}.

In astrophysics today, the interaction between components in the binaries is of
special interest, for which SySs offer an exciting laboratory \citep{b19}. For
instance, the nature of components' interaction and of the activity of WD
accretors are still controversial \citep{b34}; mass loss and stellar wind
(including the terminal velocity and acceleration mechanism) from a cool giant
are not clear; SySs have been frequently discussed as Type Ia supernovae
progenitors \citep{b50,b31,b20,b60,b16,b5}, but their contribution to the rate
of these events remains uncertain.

Theoretical studies on the formation and evolution of the SySs have been
published, e.g., by YLTK, \citet{b6,b16,b12}. Their investigations reproduced
successfully many observed properties of these objects. However, in the recent
years, a number of papers have appeared that provide new insights in the
different aspects of stellar evolution relevant to the SySs. \citet{b44} and
\citet{b57} did a more detailed study of Novae models; \citet{b36} and
\citet{b37} put forward an alternative algorithm for angular momentum loss
during the common-envelope stages of evolution; Winters and his collaborators
\citep{b53,b54,b55} studied the hydrodynamical structure of the stellar wind
around AGB stars with low mass-loss rate and low wind outflow velocity. New
observational catalogue and an analysis of SySs have been given in \citet{b61}
and \citet{b29}. It is time for a new study of the SySs.

In the present work we model the subpopulation of SySs with WDs as the hot
components and FGB or AGB stars as the cool components. We study evolutionary
channels leading to the formation of symbiotic systems in which He, CO or ONe
WDs accrete hydrogen-rich matter from the stellar wind of FGB or AGB stars. We
obtain SySs' birthrate, lifetime and number in the Galaxy, and some potentially
observable parameters of SySs, such as orbital periods, masses of the
components, their luminosities and  mass-loss rates. Special attention is paid
to the dependence of population model on the parameters entering population
synthesis. This paper is the first one of a series of papers on SySs. It offers
a basis for further studies of symbiotic stars, such as D-SySs which are SySs
with thick dust shells (paper in preparation).

In $\S$ 2 we present  our assumptions and describe some details of the modeling
algorithm. In $\S$ 3 we discuss the main results and the effects of different
parameters. In $\S$ 4 the main conclusions are given.

\section{MODELS}

In SySs, the cool component loses matter at a high rate by stellar wind and the
hot component moves in the wind and accretes enough matter to produce symbiotic
phenomenon. In this paper, binaries are considered as SySs if they satisfy the
following conditions: (i) The systems are detached; (ii) The luminosity of the
hot component is larger than 10$L_\odot$ which is the ``threshold'' luminosity
for the hot component of SySs  as inferred by \citet{b28}, \citet{b32} and
YLTK; it may be due to the thermonuclear burning (including novae outbursts,
stationary burning and post-eruption burning) or the liberation of
gravitational energy  by the accreted matter; (iii) The hot component is a WD
and the cool component is a FGB or an AGB star. Below we describe our selection
algorithm and give some details that are important for understanding the model.
All estimates made in the paper are for stars with X=0.7 and Z=0.02.
Chandrasekhar mass limit is taken as 1.44 $M_\odot$.

\subsection{Formation Channels of Symbiotic Stars}

\begin{figure}
\includegraphics[totalheight=2.5in,width=3.4in,angle=0]{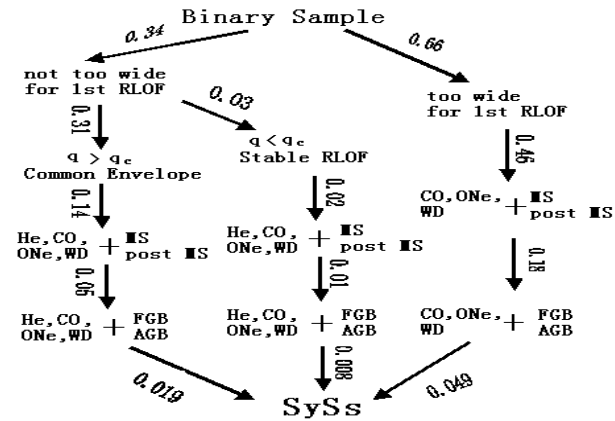}
 \caption{---Evolutionary channels for formation of SySs.
             From left  to right channels I, II and III, are shown.
             RLOF stands for the Roche lobe overflow, MS
             for main sequence, WD for white
             dwarf, FGB for the first giant branch, AGB for
             asymptotic giant branch, SySs
             for symbiotic stars.
             Numbers represent fractions of the initial sample
             of binaries that proceed through the
             consecutive evolutionary stages of each channel
             for the ``standard'' case 1 (see Table \ref{tab:results}).
             }
          \label{channel}
\end{figure}

All progenitors of SySs pass through one of the three routes
(YLTK): (i) unstable Roche lobe overflow (RLOF) with
formation of a common envelope; (ii) stable RLOF;
(iii) formation of a white dwarf+giant pair without RLOF.
These scenarios are shown in Fig.~\ref{channel}.

Through channel I pass the systems with short orbital periods. The primary
overflows its Roche lobe in the FGB or AGB stage and forms a common envelope.
After ejecting the common envelope, it transforms into a WD.

In channel II, the systems undergo stable RLOF. In this channel, the
formation of a helium star is possible. If the helium star is hot
enough and the matter inflow rate into the circum-binary medium is
high enough, then symbiotic phenomenon can arise. However, the
number of systems with helium stars is very small compared to the
number of systems with white dwarfs \citep{b60,b12} and we omit them
from consideration.

Through channel III go the systems that are initially wide. In these systems
only CO WD or ONe WD accretors are formed.

We consider ONe WDs as remnants of AGB stars that still experience
non-degenerate carbon ignition, but avoid electron captures on Ne
and Mg in the core. We assume after \citet{b42} that for solar
metallicity stars the corresponding range of masses is 6.1 -- 7.9
$M_\odot$. However, the range of masses of ONe WD progenitors still
remains uncertain and it may be shifted to higher masses and be much
more narrow, several $0.1\,M_\odot$ only \citep{it85,l4,l5,l6,l16}.
It is possible that  we overestimate the number of systems with ONe
WD {\bf and, more generally, initial-final mass relation of
\citet{b42} is too steep}.

\subsection{Common Envelope Evolution}
In channels I and II, the primary can overflow its Roche lobe. If
the mass ratio of the components ($q=M_{{\rm donor}}/M_{{\rm
accretor}}$) at the onset of RLOF is larger than a certain
critical value $q_{\rm c}$, the mass transfer is dynamically
unstable and results in the formation of a common envelope.  The
issue of the criterion for dynamically unstable RLOF $q_{\rm c}$
is still open. \citet{l7} did a detailed study
of stability of mass transfer using polytropic models.
\citet{b101,b102} showed that $q_{\rm c}$ depends heavily on the
assumed mass-transfer efficiency. In this work, we adopt $q_{\rm
c}$ after \citet{b12}:
\begin{equation}
q_{\rm c}=\left[1.67-x+2\left(\frac{M_{\rm
c}}{M}\right)^5\right]/2.13, \label{eq:qcrit}
\end{equation}
where $M_{\rm c}$ and $M$ are core mass and total mass of the
donor, respectively and  $x=0.3$ is the exponent of the
mass-radius relation at constant luminosity for giant stars
\citep{b11}.

For the common envelope evolution, it is generally assumed that the
orbital energy of the binary is used to expel the envelope of the
donor with an efficiency $\alpha_{\rm ce}$:
\begin{equation}
 E_{\rm bind}=\alpha_{\rm ce}\Delta E_{\rm orb},
 \label{eq:alpha}
\end{equation}
where $E_{\rm bind}$ is the total binding energy of the envelope and
$\Delta E_{\rm orb}$ is the orbital energy released in the
spiral-in. In the present paper we apply Eq. (\ref{eq:alpha}) in the
form  suggested by \citet{w84} with modifications after \citet{l8}:
\begin{equation}
\frac{G (M_c+M_e) M_e}{\lambda R_1} = \alpha_{ce}\left(\frac{G M_c
m}{2 a_f} -
 \frac{G M m}{2 a_i}\right).
  \label{eq:dek0}
\end{equation}
Here $\lambda$ is a structure parameter that depends on the
evolutionary stage of the donor, $a_{\rm i}$ is the orbital
separation at the onset of the common envelope, $M$, $M_{\rm c}$,
and $M_{\rm e}$ are the masses of the donor, donor's envelope and
the core,  respectively, $R_1 $ is donor's radius and $m$ is the
companion mass. Then the orbital separation of a binary after common
envelope phase $a_{\rm f}$ is given by
\begin{equation}
\frac{a_{\rm f}}{a_{\rm i}}=\frac {M_{\rm c}}{M}
\left(1+\frac{2M_{\rm e}a_{\rm i}} {\alpha_{\rm ce}\lambda m
R_{1}}\right)^{-1}. \label{eq:dek}
\end{equation}

\citet{b36} suggested to describe the variation of the separation of
components in the common envelopes by an algorithm based on the
equation for the system orbital angular momentum balance which
implicitly assumes the conservation of energy:
\begin{equation}
\frac{\Delta J}{J}=\gamma\frac{M_{\rm e}}{M+m}, \label{eq:gamma}
\end{equation}
where $J$ is the total angular momentum and $\Delta J$ is the change
of the total angular momentum during common envelope phase. The
orbital separation $a_{\rm f}$ after common envelope phase is then
\begin{equation}
\frac{a_{\rm f}}{a_{\rm i}}=\left(\frac{M}{M_{\rm
c}}\right)^2\left(\frac{M_{\rm
c}+m}{M+m}\right)\left(1-\gamma\frac{M_{\rm e}}{M+m}\right)^2.
\label{eq:nel}
\end{equation}
In the runs computed with Eq.~(\ref{eq:nel}) we apply it
irrespective of the mass ratio of components, as suggested by
\citet{b37}.

Following \citet{b37} we call the formalism of Eq. (\ref{eq:alpha})
$\alpha$-algorithm and that of Eq. (\ref{eq:gamma})
$\gamma$-algorithm. For $\alpha$-algorithm, there are two
parameters: $\alpha_{\rm ce}$ and $\lambda$. Both parameters are
highly uncertain. It's still not completely clear whether sources
other than gravitational energy have to be taken into account when
computing $\alpha_{\rm ce}$ and $\lambda$ and how the core-envelope
boundary for the estimate of $\lambda$ has to be defined \citep[see,
e. g.,][]{b7,b15,dt00,td01,sh03}. Both parameters depend on the
evolutionary stage in which RLOF occurs. In the absence of
prescriptions for determination of $\alpha_{\rm ce}$ and $\lambda$
for particular donor+accretor combinations one can consider  the
``combined'' parameter $\alpha_{\rm ce}\lambda$ only. We make
computations for $\alpha_{\rm ce}\lambda$=0.5, 1.5, and 2.5. The
choice of $\alpha_{\rm ce}\lambda >1$ is also motivated by the fact
that YLTK considered a formulation of the common envelope equation
which gives for $a_{\rm f}/a_{\rm i}$ values that are different from
the ones given by Eq. (\ref{eq:dek}) and is equivalent to the usage
of $\alpha_{\rm ce}\lambda >1$ in Eq.~(\ref{eq:dek}). Thus, we test
both the influence of the uncertainties in the definitions of
$\alpha_{\rm ce}$ and $\lambda$ and in the formulation of the
common-envelope equation.

Generally, we make numerical simulations for the $\alpha$-algorithm,
but we also check the $\gamma$-algorithm effects on the formation of
SySs, see Tables \ref{tab:cases} and \ref{tab:results}.  For
$\gamma$, we consider two values -- 1.5 and 1.75 \citep{b37}. For
both algorithms we assume that companion mass does not change during
common envelope phase since this stage is very short.

\subsection{The Model of Symbiotic Stars }

In SySs, the cool component is in FGB or AGB stage. Because in the core He
burning stage the radius and luminosity of the cool component are lower and it
can offer very little mass by stellar wind for the hot component accretor, this
stage does not contribute to the total population of SySs \citep{b60}. We
therefore skip this stage in our work.

\subsubsection{Mass Loss}
No comprehensive theory of mass loss for AGB stars exists at
present. In this paper, we accept the prescription of \citet{b11}.
In MS, Hertzsprung Gap and FGB stages, we apply
\citet{b45} mass-loss law
\begin{equation}
\dot{M}(M_\odot {\rm yr}^{-1})=4\times 10^{-13}\frac{\eta
LR}{M}, \label{eq:RMML}
\end{equation}
where $\eta=0.5$ and $L, R$ and $M$  are the luminosity, radius
and mass of the star in solar units, respectively. In AGB stage,
we use the mass-loss law suggested by \citet{b51}
\begin{equation}
\log_{10} (\dot{M})= -11.4+0.0123(P-100 \max(M/M_\odot-2.5,
0.0)),~ \label{eq:agbml}
\end{equation}
where $P$ is the Mira pulsation period in days given by
\begin{equation}
\log_{10}(P) =-2.07+1.94 \log_{10}(R/R_\odot)-0.90
\log_{10}(M/M_\odot).
\end{equation}
When $P \geq 500$ days, the steady super-wind phase is modeled by
the law
\begin{equation}
\dot{M}(M_\odot {\rm yr}^{-1})=2.06\times 10^{-8}\frac{L/L_\odot}{v_\infty},
\end{equation}
where $v_\infty$ is the terminal velocity of the super-wind in km
s$^{-1}$; we use $v_\infty$=15 km s$^{-1}$ in this paper.

If a star loses mass by stellar wind, it loses angular momentum
too. We assume that the lost matter takes away the specific
angular momentum of the prospective donor. In this work, the tidal
enhancement of mass loss \citep{b49} is not considered.

For giant-star accretors we assume that the accretion rate is
unlimited, since the stars with deep convective envelopes may
shrink in the dynamical time scale in response to the mass
increase.  For non-WD accretors  accretion rate is limited by the
thermal time scale: $\dot{M}_2 \le M_2 / \tau_{\rm th,2}$. For
larger $\dot{M}_2$, a common envelope forms.

\subsubsection{Accretion Rate of Stellar Wind}
Stellar wind accretion is crucial for the SySs phenomenon.
 \citet{b47} and \citet{b25} showed that for
SySs the classical \citet{b1} accretion formula for the stellar
wind is generally valid. The mean accretion rate is:
\begin{equation}
\dot{M}_{{\rm hot}}=\frac{-1}{\sqrt{1-e^2}}\left(\frac{GM_{{\rm
hot}}}{v^2_{\rm w}}\right)^2\frac{\xi_{\rm w}}{2a^2}
\frac{1}{(1+v^2)^{3/2}}\dot{M}_{{\rm cool}}, \label{eq:bh}
\end{equation}
where $1\le\xi_{\rm w}\le2$ is a parameter ($\xi_{\rm
w}=\frac{3}{2}$ in this work), $v_w$ is the wind velocity and
\begin{equation}
v^2=\frac{v^2_{{\rm orb}}}{v^2_{\rm w}},~~
v^2_{{\rm
orb}}=\frac{GM_{\rm t}}{a},
\end{equation}
where $a$ is the semi-major axis of the ellipse, $v_{{\rm orb}}$
is the orbital velocity and total mass $M_{\rm t}=M_{{\rm
hot}}+M_{{\rm cool}}$.

If a hot component accretes from  the stellar wind of a cool
component, a fraction of the angular momentum lost by the cool
component is  transferred into the spin of the hot component.
Following \citet{b12}, we assume that the efficiency of the
angular momentum transfer is 1.

\subsubsection{Wind Velocity}
\label{subsec:wind_vel}
Accretion rate of the stellar wind
[Eq.~(\ref{eq:bh})], strongly depends on the wind velocity $v_{\rm
w}$ which is not readily determined. Taking into account the fact
that the hot component may be located in the zone of stellar wind
acceleration, we assume that the velocity of the stellar wind is
related to the terminal wind velocity $v_\infty$.

YLTK have defined the wind velocity as
\begin{equation}
v_{\rm w}=\alpha_{\rm w}v_\infty,
\label{eq:vwvinfty}
\end{equation}
where $v_\infty$ is the terminal wind velocity and $\alpha_{\rm
w}$ is approximated by
\begin{equation}
\alpha_{\rm w}=\frac{0.04(r/R_{\rm d})^2}{1+0.04(r/R_{\rm d})^2},
\label{eq:windyl}
\end{equation}
where $r$ is the distance from the donor and $R_{\rm d}$ is the
radius of the donor. \citet{b52} gave an empirical formula for the
SyS EG Andromedae. This velocity law is:
\begin{equation}
\frac{v_{\rm w}}{v_\infty}=\alpha_{\rm w}=\left\{
\begin{array}{ll}
c_1(r/R_{\rm d})^{10} & {\rm for}\ \   r/R_{\rm d}\leq 3.75\\
1-{\rm e}^{-c_2(r/R_{\rm d}-c_3)} & {\rm for} \ \ r/R_{\rm d}> 3.75,\\
\end{array}
\right. \label{eq:windvo}
\end{equation}
where $c_1, c_2$ and $c_3$ are $0.2\times(3.75)^{-10},
\frac{2}{3}$ and 3.42. Below we present results of numerical
simulations for the values of $\alpha_{\rm w}$ given by Eqs.
(\ref{eq:windyl}), (\ref{eq:windvo}) and for $\alpha_{\rm w}=1$.

 The definition of terminal velocity in the literature is not
unique. YLTK take $v_\infty$ as $v_{\rm esc}$ which is the
surface escape velocity. But, as noted by  \citet{b10}, the main
characteristic of  evolved K and early M-giants' cool winds is
that their terminal velocities are lower than the surface escape
velocity, typically, $v_\infty\leq\frac{1}{2}v_{{\rm esc}}$.
\citet{b53} found two dynamically different regimes for the
spherical outflows of AGB stars: in the first regime, the terminal
wind velocity is in excess of 5 km s$^{-1}$ and the mass-loss rate
is higher than $3\times 10^{-7} M_\odot$yr$^{-1}$; in the second
regime, the terminal wind velocity is smaller than 5 km s$^{-1}$
and the mass-loss rate is lower than $3\times 10^{-7}
M_\odot$yr$^{-1}$. \citet{b55} fitted the relation between
mass-loss rates and terminal wind velocities derived from their
CO(2-1) observations by
\begin{equation}
\log_{10} (\dot{M}[M_\odot{\rm yr}^{-1}])=-7.40+\frac{4}{3}\log_{10}
(v_{\infty}/{\rm km \, s^{-1}}). \label{eq:winters}
\end{equation}

This relation is very close to the results of \citet{b39}.
Mass-loss rates can be obtained by Eq. (\ref{eq:agbml}) and
$v_\infty$ can be calculated from Eq.  (\ref{eq:winters}).
\citet{b53} showed the radial structure of hydrodynamic velocity
for low and high mass-loss rate models. Obviously, Eqs.
(\ref{eq:windyl}) or (\ref{eq:windvo}), which predict the increase
of outflow velocity, are not adequate for the low mass-loss rate
model.
 Semi-regular variable L$^2$ Pup, whose
mass-loss rate is lower than $3\times 10^{-7}M_\odot$yr$^{-1}$
\citep{b54}, is an example: its photospheric material is moving at
a velocity of about 10 km s$^{-1}$ which is much higher than its
terminal wind velocity of several km s$^{-1}$. In the absence of a
clear theory of stellar wind generation and behaviour for low-mass
giants we make numerical simulations
for a range of assumptions on stellar wind. \\
For FGB stars we consider two cases:\\
(a) $v_{\infty}=\frac{1}{2}v_{\rm esc}$;\\
(b) $v_{\infty}=v_{\rm esc}$.\\
For AGB stars we consider cases (a),(b),
and case \\
(c) in which $v_{\infty}$ is determined as follows:\\
 For mass loss rate
higher than $3\times10^{-7}M_\odot$yr$^{-1}$, $v_\infty$ is
determined by Eq.~(\ref{eq:winters}). However,
Eq.~(\ref{eq:winters}) is valid for $\dot{M}$ close to $
10^{-6}M_\odot$ yr$^{-1}$. For higher mass loss rate,
Eq.~(\ref{eq:winters}) gives too high a $v_\infty$. Based on the
models of \citet{b53}, we assume $v_\infty=\min(30 {\rm km\,
s^{-1}}, v_\infty)$. Wind velocity is given by
Eq.~(\ref{eq:vwvinfty}).
where $\alpha_{\rm w}$ is defined by Eq.~(\ref{eq:windyl}).\\
 For
$\dot{M} \leq 3.0 \times 10^{-7}M_\odot$yr$^{-1}$, we assume that wind
velocity decelerates from $v_{\rm esc}$ at the stellar surface to
5 km s$^{-1}$ at $r/R_{\rm d}=10$,  using an {\it ad hoc} function:
\begin{equation}
v_{\rm w}=\left\{
\begin{array}{lll}
\frac{5{\rm km\, s}^{-1}-v_{\rm esc}}{9}(r/R_{\rm d})+\frac{10v_{\rm esc}-5{\rm km\, s}^{-1}}
{9}& \ \ \ \ r\leq 10R_{\rm d}\\
5 {\rm km\, s}^{-1}&\ \ \ \ r>10R_{\rm d}.
\end{array}
\right.
\end{equation}
(d) In addition, a model with the ``standard'' terminal wind
velocity of AGB stars equal to 15 km s$^{-1}$ is calculated.

\subsubsection{Critical Mass-Accretion Rate of WD }
\label{sec:critical} Most models for eruptions in SySs with WD
accretors involve the nuclear burning of the material accreted by
WDs \citep{b28}. Based on \citet{b50} and \citet{b40}, SySs powered
by hydrogen burning on the surface of WDs can be
 divided further into two subgroups: ``ordinary'' SySs, which
are assumed to burn hydrogen steadily if the mass-accretion rate
is above a critical rate, $\dot{M}_{{\rm st}}$, and symbiotic
novae (SyNe)\footnote{SyNe represents symbiotic nebulae in some
literature but symbiotic novae in this paper.}, which experience
thermonuclear runaways in their surface hydrogen layers if the
mass-accretion rate is lower than $\dot{M}_{{\rm st}}$. In SyNe
stage and stable hydrogen burning stage, the luminosity of the hot
component due to hydrogen burning is given by the following
approximation for the core mass-luminosity relation for cold WDs
accreting hydrogen \citep{b16}:
\begin{equation}
L/L_\odot\approx 4.6\times 10^4(M_{{\rm core}}/M_\odot -0.26).
\label{eq:mcl}
\end{equation}

For $\dot{M}_{{\rm st}}$, following YLTK, we use the
approximation to the results of \citet{b13} given by
\begin{equation}
\log_{10} \dot{M}_{{\rm st}}(M_\odot {\rm
yr}^{-1})=-9.31+4.12M_{\rm WD}-1.42(M_{\rm WD})^2,
\label{eq:steady}
\end{equation}
with $M_{\rm WD}$ in solar units. For steady burning, if the
mass-accretion rate is above a certain critical rate,
$\dot{M}_{{\rm cr}}$, an accreting WD may evolve into a giant
\citep{b19} or generate an optically thick wind \citep{b62}. Based
on the results of \citet{b13}, an approximation to $\dot{M}_{{\rm
cr}}$ is given by
\begin{equation}
\begin{array}{ll}
 \log_{10} \dot{M}_{{\rm cr}}(M_\odot {\rm
yr}^{-1})=&-9.78+9.16M_{\rm
WD}-8.13(M_{\rm
WD})^2\\
 & +2.44(M_{\rm WD})^3,\\
\end{array}
\end{equation}
where $M_{\rm WD}$ is in $M_\odot$. If the mass-accretion rate is
higher than $\dot{M}_{{\rm cr}}$, we assume that the accreted
hydrogen burns steadily at the surface of the WD and hydrogen-rich
matter is converted into helium at the rate  given by Eq.
(\ref{eq:alphah}) below, while the unprocessed matter is lost from
the system as an optically thick wind and takes away specific
angular momentum of the accretor. We also made a test run (case 13
in Table \ref{tab:cases}) assuming that excess matter expands and
transforms the WD into a giant.

In addition, if the accretion rate is higher than a certain value,
the release of gravitational energy ``mimics'' steady hydrogen
burning.
If the luminosity of the WD due to the liberated gravitational
potential energy, $L_{{\rm grav}}$, is larger than 10$L_\odot$, we
assumed that the system powered by accretion is in the ``symbiotic
stage'' too. $L_{{\rm grav}}$\ (in solar units) is given by
\begin{equation}
L_{{\rm grav}}\sim 3.1\times 10^7 \dot{M}_{{\rm acc}}\frac{M_{{\rm
WD}}}{R_{{\rm WD}}},
\label{eq:lgrav}
\end{equation}
where $\dot{M}_{{\rm acc}}$ is in $M_\odot$yr$^{-1}$, and $M_{\rm
WD}$ and $R_{\rm WD}$ are the mass and  radius of an accreting WD
in solar units, respectively. We note also that variations in
mass-flow rate may cause optical outbursts that have distinct
spectral features \citep[see discussion in ][]{b28}.

\subsubsection{Critical Ignition Mass}

 For SyNe a certain amount of matter has to be accumulated prior
to the first explosion. Following
YLTK we use the ``constant pressure'' expression for $\Delta
M_{\rm crit}$ which implies that the ignition occurs when the
pressure at the base of the accreted layer rises to a certain
limit:
\begin{equation}
\frac{\Delta M_{{\rm crit}}^{{\rm WD}}}{M_\odot}=2\times
10^{-6}\left(\frac{M_{{\rm WD}}}{R^4_{{\rm WD}}}\right)^{-0.8},
\label{eq:yunm}
\end{equation}
where $R_{{\rm WD}}$ is the radius of zero-temperature degenerate
objects \citep{b35}:
\begin{equation}
R_{{\rm WD}}=0.0112R_\odot[(M_{{\rm WD}}/M_{{\rm
ch}})^{-2/3}-(M_{{\rm WD}}/M_{{\rm ch}})^{2/3}]^{1/2}
\end{equation}
with $M_{{\rm ch}}=1.433M_{\odot}$ and $R_{\odot}=7\times 10^{10}$
cm. In fact $\Delta M_{\rm crit}^{\rm WD}$ is a highly sensitive
function of the temperature, mass, and accretion rate of a WD, as
well as of assumptions on the nature of the mixing process at the
base of the accreted layer and actual nuclear abundances in the
accretor. Detailed grids of nova models for CO WD-accretors may be
found in \citet{b44} and \citet{b57}. Eq. (\ref{eq:yunm}) agrees
with the results of numerical calculations for relatively cold WDs
with the temperature $10^7$~K  \citep{b57} within a factor of 5,
except for the model with a 1.4$M_\odot$ WD and $\dot{m}=10^{-7}
M_\odot$ yr$^{-1}$, for which agreement is within a factor of 7.
 \citet{b38}
gave numerical fits to the critical ignition masses for novae
models calculated by \citet{b44}. We made a run of the code  using
 the fitting formula for relatively cold ($T\sim 10^7$K) WD
 [see Eq. (A1) of \citet{b38} in
which $\Delta M_{\rm crit}^{\rm WD}$ depends on the mass accretion
rates and masses of WD accretors]. Its extrapolation to lower mass
(0.4 $M_\odot$) agrees with the results of numerical calculations
[by the same code; \citet{b57}] to within a factor less than 3.
But for the lowest mass-accretion rates ($10^{-12}, 10^{-12.3}
M_\odot$yr$^{-1}$), its errors are larger. Since the systems with
such low mass-accretion rates provide only a tiny contribution to
the total number of SySs, we use fits for
$10^{-11}M_\odot$yr$^{-1}$ in this case. Results agree with the
ones in \citet{b57} to within a factor of less than 3.

 Because of the complicated
dependence of $\Delta M_{\rm crit}$ on the input parameters we ran
several simulations varying $\Delta M_{\rm crit}$, see
Table~\ref{tab:cases}.

Oxygen-neon WDs need to accrete a more massive envelope than the
same mass CO WDs before the outbursts begin \citep{l14}, because
of the lower $^{12}$C abundance in the accretor. Using \citet{l14}
data,
we roughly assume that $\Delta M_{\rm crit}$ for an ONe WD is
twice that for the same mass CO WD.

\subsubsection{Mass-Accumulation Efficiency}
\label{maeff}
The efficiency of mass accumulation by a WD can
never be 100\%. First, it strongly depends on the strength of
symbiotic novae and, second, even steady-burning WDs blow stellar
wind. We define the ratio  of the mass of burnt hydrogen to the
mass of the matter accreted by a WD as $\alpha_{\rm H}$. The strength
of a symbiotic nova depends on the mass and mass-accretion rate of
the WD. According to \citet{b57}, we roughly define the boundary
between strong SyNe and weak SyNe by the mass-accretion rate $\dot
{M}_{\rm ws}$  (in $M_\odot {\rm
yr}^{-1}$)
which is given by
\begin{equation}
\log_{10}\dot{M}_{\rm ws}=\left \{
\begin{array}{ll}
-11.01+6M_{\rm WD}& \\
-1.90M_{\rm WD}^2, &\ \ {\rm for}\ M_{\rm WD}\leq 1M_\odot;\\
-7.0,&\ \ \ {\rm for}\ M_{\rm WD}> 1M_\odot .\\
\end{array}
\right.
\end{equation}

For weak SyNe, \textit{i.e.}, mass-accretion rates being between
$\dot{M}_{\rm st}$ and $\dot{M}_{\rm ws}$, we use an approximation
to $\alpha_{\rm H}$ based on Fig. 2 in YLTK [also Fig. 16
of \citet{b16}]:
\begin{equation}
\alpha_{\rm H}=\left\{
\begin{array}{lll}
-4.39-1.48\log_{10}\dot{M}_{\rm acc}&
\\-0.10(\log_{10}\dot{M}_{\rm acc})^2, & {\rm for}\ \ \log_{10} \dot{M}_{\rm acc}<-6.36;\\
11.66+4.56\log_{10} \dot{M}_{\rm acc} & \\
+0.45(\log_{10}\dot{M}_{\rm acc})^2,& {\rm for}\ \ \log_{10} \dot{M}_{\rm acc}\geq-6.36\\
\end{array}
\right. \label{eq:alphah}
\end{equation}
 where $\dot{M}_{\rm acc}$ is in $M_\odot$yr$^{-1}$.

For strong SyNe, \textit{ i.e.}, mass-accretion rates lower than
$\dot{M}_{\rm ws}$, most of the accreted mater is expelled and in
some cases even an erosion of the WD occurs \citep{b44,b57}. Using
data on the ejected mass, $M_{\rm ej}$, helium mass fraction in
the ejecta, $Y_{\rm ej}$, and in the convective envelope, $Y_{\rm
env}$, heavy element mass fraction in the envelope, $Z_{\rm ej}$,
and in the ejecta, $Z_{\rm ej}$, given in Table 2 of \citet{b57},
the
 mass of the burnt hydrogen can be roughly calculated as:\\
for $\Delta M_{\rm crit}\geq M_{\rm ej}$,
\begin{equation}
\begin{array}{ll}
\Delta M_{\rm H}=& \Delta M_{\rm crit}\times X_{\rm H}-M_{\rm
ej}(1-Y_{\rm ej}-Z_{\rm ej})\\
 &-(\Delta M_{\rm crit}-M_{\rm ej})(1-Y_{\rm
env}-Z_{\rm env}),\\
\end{array}
\end{equation}
where we assume that the mixing of the accreted layer with core
material does not occur;\\
for $\Delta M_{\rm crit}< M_{\rm ej}$,
the amount of hydrogen burnt during an outburst is approximated by
the difference between $Y_{\rm env}$ and $Y_{\rm ej}$ which may be
computed using data from \citet{b44},
\begin{equation}
\Delta M_{\rm H}= \Delta M_{\rm crit}Y_{\rm env}-M_{\rm ej}Y_{\rm ej}.
\end{equation}

Then, using data from \citet{b57}, we fit $\alpha_{\rm H}$ by
\begin{equation}
\begin{array}{ll}
\alpha_{\rm H} &
=\frac{\Delta M_{\rm H}}{\Delta M_{\rm crit}}\\
 &=-0.1391+0.7548M_{\rm WD}\\
  &-1.0124M_{\rm WD}^2+0.4739M_{\rm WD}^3,\\
\label{eq:fitalpha}
\end{array}
\end{equation}
with $M_{\rm WD}$ in $M_\odot$.

For $\alpha_{\rm H}$, if the difference between $M_{\rm crit}$ and
$M_{\rm ej}$ is neglected and the greatest part of the energy
released 
in the  outburst is used to lift the ejected shell from the
gravitational potential well of the WD, then, according to \citet{b57},
we can give a rough analytical estimate:
\begin{equation}
\alpha_{\rm H} \approx 4.54\times10^{-4}\frac{M_{\rm WD}}{R_{\rm WD}},
\label{eq:yaron}
\end{equation}
where $M_{\rm WD}$ is the mass of the WD accretor and $R_{\rm WD}$
is its radius in solar units. $\alpha_{\rm H}$ in
Eq.~(\ref{eq:fitalpha}) agrees with that in Eq.~(\ref{eq:yaron})
within a factor of 3 for $0.4M_\odot\leq M_{\rm WD}\leq
1.4M_\odot$ and the former is usually larger than the latter for a
given $M_{\rm WD}$.

After a nova, SySs spend  certain time in the ``plateau'' stage
with high luminosity. This stage lasts for
\begin{equation}
t_{\rm on}=6.9\times 10^{10}\frac{\alpha_{\rm H}\Delta M_{\rm
crit}^{\rm WD}}{L}~{\rm yr}, \label{eq:t_on}
\end{equation}
where $L$ is given by Eq.~(\ref{eq:mcl}). Stably burning hydrogen
``ordinary'' SySs are always in the ``plateau'' stage, their
$\alpha_{\rm H}$ is given by Eq.~(\ref{eq:alphah}). In this work the
stage of hydrogen burning is denoted as ``on''-stage.

For weak SyNe and stable hydrogen burning, we assume that the
burnt hydrogen is deposited at the surface of the WD and that the
unprocessed matter is lost from the system by the stellar wind  that
takes away specific angular momentum of the WD.
 For strong SyNe, the mass of the WD increases if $M_{\rm
ej}<\Delta M_{\rm crit}$. If $M_{\rm ej}>\Delta M_{\rm crit}$, the
WD is eroded. We fit the \citet{b57} data by a formula
\begin{equation}
\begin{array}{ll}
\frac{M_{\rm ej}}{\Delta M_{\rm crit}}=&
-28.4700-9.6207\log_{10}\dot{M}_{\rm acc}\\
 & -1.0372(\log_{10}\dot{M}_{\rm
acc})^2-0.0371(\log_{10}\dot{M}_{\rm acc})^3,\\
\end{array}
\end{equation}
where $\dot{M}_{\rm acc}$ is in $M_\odot$yr$^{-1}$ and which agrees
with numerical results to within a factor of 1.4.

\subsubsection{Decline of SyNe}

After the outburst, the system remains observable as a SyS for a
time span $t_{\rm cool}$, until the WD cools to the temperature
at which its luminosity becomes lower than 10$L_\odot$.
\citet{b43} showed that the luminosity declines with time as
$L\propto t^{-1.14}$. \citet{b48} confirmed that the observed
cooling rate of the WD is consistent with the model of
\citet{b43}. Accordingly, we assume that after ``plateau" stage
the WD enters  decline stage during which its luminosity
decreases as
\begin{equation}
L(t)=L(0) t^{-1.14},
 \label{eq:decline}
\end{equation}
where $L(0)$ is given by Eq.~(\ref{eq:mcl}) and $t$ is in years.
When $L(t)=10L_\odot$, SySs stage terminates and the cooling time
$t_{\rm cool}$ can be obtained. The lifetime of a SyS is the sum
of $t_{\rm on}$ and $t_{\rm cool}$.

We assume that during the outburst WD does not accrete any matter,
due to the high-velocity wind and high luminosity. At the onset of
decline,  WD begins to accrete matter from the stellar wind of the
cool giant. At the same time, the WD loses mass due to intense
radiatively-driven stellar wind \citep{b60,b16}. We roughly assume
that the rate of the mass loss may be estimated by
\begin{equation}
\dot{M}=\frac{L}{v_{\rm esc}c}, \label{eq:hotml}
\end{equation}
where $c$ is the speed of light, $v_{\rm esc}$ is escape velocity
and $L$ is the luminosity of the WD given by
Eq.~(\ref{eq:decline}). In Eq. (\ref{eq:hotml}), $v_{\rm esc}$ is
given by
\begin{equation}
v_{\rm esc}=\left(\frac{2GM_{\rm WD}}{R_{\rm WD}}\right)^{1/2}.
\label{eq:vesc}
\end{equation}
In Eq.~(\ref{eq:vesc}), $R_{\rm WD}$ is roughly taken as the
radius of a cool WD. The mass-loss rate given by
Eq.~(\ref{eq:hotml}) is uncertain by a factor of several, since
during the main part of the ``on''-stage the actual radius of the WD
is by about an order of magnitude larger than the radius of a cool
WD \citep[see, e.g.,][]{b32,b33} and the performance number of the
radiative stellar wind, $\dot{M} v_{\rm esc}/(L/c)$, is not known.

Before the luminosity of the WD declines  to $10L_\odot$, the
decline stage may be terminated if the amount of accreted matter
is larger than the critical ignition mass.

\subsubsection{Helium Flashes}

When the helium shell accumulates up to a critical mass, the
helium flash occurs. This critical mass is about (0.1 -- 0.2)
$M_\odot$ [for initially cold and non-rotating WDs and $\dot
M\lesssim3\times10^{-8}M_\odot$yr$^{-1}$; e.g., \citet{b14,b56}].
After a hydrogen flash, the WD becomes ``hot" and the helium
flashes require ignition mass which is much lower than the
above-mentioned critical mass. For hot white dwarfs we use an
approximation for the critical mass suggested by \citet{b13,b16}
\begin{equation}
\Delta M_{\rm crit}^{\rm He}=10^{6.65}R^{3.75}_{\rm
WD}M^{-0.3}_{\rm WD}\dot{M}_{-8}^{-0.57}, \label{eq:Hemc}
\end{equation}
where $\dot{M}_{-8}$ is the accretion rate in units of
$10^{-8}M_\odot$yr$^{-1}$ and the remaining variables  are in
solar units. But Eq. (\ref{eq:Hemc}) is not valid for the helium
accumulation rate lower than $10^{-8}M_\odot$yr$^{-1}$. We take
the critical mass for helium ignition as 0.1 $M_\odot$ when the
helium accumulation rate is lower than $10^{-8}M_\odot$yr$^{-1}$.
Mass accumulation efficiency in the helium shell flashes, for a
wide grid of masses and accretion rates, is computed by
\citet{b17}. We find the growth rate of white dwarfs by linear
interpolation in these grids. We should note, however, that if
white-dwarf accretors are rapid rotators, the helium accumulation
efficiency at low $\dot{M}$ may strongly reduce due to the
dissipation of energy at the base of the accreted layer
\citep{b82}.

\subsection{Conversion to Observed Parameters}
To compare our results with observations, we convert
luminosities of cool components into absolute magnitudes
using bolometric corrections which can be obtained by
interpolation in the BaSeL-2.0 stellar spectra library of
\citet{b22,b23}.
SySs are usually detected due to the presence of
features typical for ionized nebulae in the spectra of cool
components. This allows us to assume that the simplest selection
effect that governs the observed sample of SySs is the visual
magnitude of the cool components, $V_{\rm c}$ \citep{b60}.
Inspection of the catalogue of SySs \citep{b61} reveals that the
number of objects per unit interval of stellar magnitude increases
up to $V_{\rm c}\sim 12.0$ mag. In both \citet{b19} and
\citet{b61}, there are about 50 SySs with $V_{\rm c}\leq 12.0$
mag. So, we take 12.0 mag as the limiting stellar magnitude of the
complete sample and compute the number of model objects with
$V_{\rm c}\leq 12.0$ mag.

The number of stars within the distance $d$ from the Sun is taken to
be \citep{b3,b6}
\begin{equation}
N(d,t)=\frac{1}{d_0^2t_{\rm Gal}}\frac{d^3}{\sqrt{d^2+h^2}},
\end{equation}
where $t$ is the evolutionary age of the star, $h=1200(t/t_{{\rm
Gal}})^{1/2}$ and $t_{{\rm Gal}}$ is the Galactic age. The
quantity $d_0$ is defined such that $1/\pi d_0^2$ is the density
per unit area in the Galactic plane of stars in the solar
neighborhood projected onto the plane; following \citet{b3}, we
take $d_0=0.054$ pc and $t_{\rm Gal}$=12Gyr. \citet{b3} used the
distance $d\leq 3$ kpc, we roughly extrapolate $d$ to 15 kpc in
this work for comparison with observations.

To estimate interstellar extinction, we use the corrected relation
from \citet{l17}
\begin{equation}
A_{\rm V}=0.14\times  \csc |b| [1-{\rm \exp}{(-10 d \sin
|b|)}]~ {\rm mag.},
 \label{eq:extin}
\end{equation}
where $b$ is the galactic latitude and $d$ the distance in kpc.
\citet{b19} showed that Galactic SySs strongly concentrate toward
the plane. We roughly take stellar extinction as 1.4 mag per kpc
which is the maximum value based on Eq. (\ref{eq:extin}).

\subsection{Basic Parameters of the Monte Carlo Simulation }

For the  population synthesis for binary stars, the main input model
parameters are: (i) the initial mass function (IMF) of the
primaries; (ii) the mass-ratio distribution of the binaries; (iii)
 the distribution of orbital separations; (iv) the eccentricity
distribution; (v) the metallicity $Z$ of the binary systems.

A simple approximation to the IMF of \citet{b30} is used. The
primary mass is generated using  the formula suggested by
\citet{b3}
\begin{equation}
M_1=\frac{0.19X}{(1-X)^{0.75}+0.032(1-X)^{0.25}},
\end{equation}
where $X$ is a random variable uniformly distributed in the range
[0,1],  and $M_1$ is the primary mass from $0.8M_\odot$ to
$8M_\odot$.

The mass-ratio distribution is quite controversial. We consider
only a constant mass-ratio distribution \citep{b27,b4},
\begin{equation}
n(q)=1,~~    0< q \leq 1,
\end{equation}
where $q=M_2/M_1$.

The distribution of separations is given by
\begin{equation}
\log a =5X+1,
\end{equation}
where $X$ is a random variable uniformly distributed in the range
[0,1] and $a$ is in $R_\odot$.

We assume that all binaries have initially circular orbits. We
follow the evolution of both components by the rapid binary
evolution code, including the effect of tides on binary evolution
\citep{b12}. We take $2\times10^5$ initial binary systems for each
simulation. Since we present, for every simulation, the results of
one run of the code, the numbers given are subject to Poisson noise.
For simulations with $2\times10^5$ binaries, the relative errors of
the numbers of symbiotic systems of different kinds are lower than
7\%, with exception of SySs with He WD accretors in cases 6, 10 and
11 (see Table~\ref{tab:results}). Under model assumptions for cases
6, 10 and 11, SySs with He WD accretors hardly form and the numbers
of them expected in the Galaxy are so small (much less than 1 see
Table~\ref{tab:results}) that it is impossible to observe them. We
have made a control run for case 10 with a sample of $6\times 10^5$
binary systems. The maximum difference among the results obtained
for $2\times10^5$ and $6\times 10^5$ samples is less than 7\% of the
numbers presented in Table~\ref{tab:results}, except for the results
on SySs with He WD accretors which agree with the numbers given in
columns 2, 10, 14 and 22 of Table~\ref{tab:results} within a factor
about 2. However, systems with He-accretors constitute only  a minor
fraction of the total SySs population in case 10 and, hence,
inaccuracy in their number has a negligible effect upon our main
results. Thus, $2\times10^5$ initial binaries appear to be an
acceptable sample for our study.

To
calculate the birthrate of SySs, we assume that one binary with
$M_1\geq 0.8 M_\odot $ is formed annually in the Galaxy
\citep{b58,b6,b60}.
\begin{table}
 \begin{minipage}{85mm}
  \caption{Parameters of the models of the population of symbiotic stars.
           Case 1$^\star$ means the standard model. In case 6,
           $v_\infty$ of FGB stars is taken as $v_{\rm esc}$ and
           it is $\frac{1}{2}v_{\rm esc}$ in other cases. In case 7, wind velocity
           $v_{\rm w}$  is treated as described under item (c) in subsection
          \ref{subsec:wind_vel}. In case 13, critical ignition mass is taken
          after Eq. (A1) from \citet{b38}. }
  \tabcolsep1.0mm
  \begin{tabular}{|llllll|}
  \hline\hline
case & Common  & $v(\infty)$ &$\alpha_{\rm w}$ &$\Delta M_{\rm crit}^{\rm WD}$& Optically \\
     & envelope&             &                 &                     & thick wind \\
\hline\hline
case 1$^\star$. & $\alpha_{\rm ce}\lambda=0.5 $ & $\frac{1}{2}v_{\rm esc}$& Eq. (13)& $\Delta M_{{\rm crit}}^{\rm WD}$             & on \\[1.0mm]
case 2. & $\alpha_{\rm ce}\lambda=1.5 $ & $\frac{1}{2}v_{\rm esc}$& Eq. (13)& $\Delta M_{{\rm crit}}^{\rm WD}$             & on \\[1.0mm]
case 3. & $\alpha_{\rm ce}\lambda=2.5 $ & $\frac{1}{2}v_{\rm esc}$& Eq. (13)& $\Delta M_{{\rm crit}}^{\rm WD}$             & on \\[1.0mm]
case 4. & $\gamma=1.5 $          & $\frac{1}{2}v_{\rm esc}$& Eq. (13)& $\Delta M_{{\rm crit}}^{\rm WD}$             & on \\[1.0mm]
case 5. & $\gamma=1.75$          & $\frac{1}{2}v_{\rm esc}$& Eq. (13)& $\Delta M_{{\rm crit}}^{\rm WD}$             & on \\[1.0mm]
case 6. & $\alpha_{\rm ce}\lambda=0.5 $ & $           v_{\rm esc}$& Eq. (13)& $\Delta M_{{\rm crit}}^{\rm WD}$             & on \\[1.0mm]
case 7. & $\alpha_{\rm ce}\lambda=0.5 $ & Eq.  $(\ref{eq:winters})$& Eq. (13)& $\Delta M_{{\rm crit}}^{\rm WD}$             & on \\[1.0mm]
case 8. & $\alpha_{\rm ce}\lambda=0.5 $ &            15km s$^{-1}$& Eq. (13)& $\Delta M_{{\rm crit}}^{\rm WD}$             & on \\[1.0mm]
case 9. & $\alpha_{\rm ce}\lambda=0.5 $ & $\frac{1}{2}v_{\rm esc}$& Eq. (14)& $\Delta M_{{\rm crit}}^{\rm WD}$             & on \\[1.0mm]
case 10.& $\alpha_{\rm ce}\lambda=0.5 $ & $\frac{1}{2}v_{\rm esc}$& 1      & $\Delta M_{{\rm crit}}^{\rm WD}$             & on \\[1.0mm]
case 11.& $\alpha_{\rm ce}\lambda=0.5 $ & $\frac{1}{2}v_{\rm esc}$& Eq. (13)& $3\Delta M_{{\rm crit}}^{\rm WD}$            & on \\[1.0mm]
case 12.& $\alpha_{\rm ce}\lambda=0.5 $ & $\frac{1}{2}v_{\rm esc}$& Eq. (13)& $\frac{1}{3}\Delta M_{{\rm crit}}^{\rm WD}$  & on \\[1.0mm]
case 13.& $\alpha_{\rm ce}\lambda=0.5 $ & $\frac{1}{2}v_{\rm esc}$& Eq. (13)& $^{\star}\Delta M_{{\rm crit}}^{\rm WD}$ & on \\[1.0mm]
case 14.& $\alpha_{\rm ce}\lambda=0.5 $ & $\frac{1}{2}v_{\rm esc}$& Eq. (13)& $\Delta M_{{\rm crit}}^{\rm WD}$             & off \\

\hline
 \label{tab:cases}
\end{tabular}
\end{minipage}
\end{table}
\section{RESULTS}

We construct a set of models in which we vary different input
parameters relevant to the symbiotic phenomenon produced by
hydrogen burning at the surface of WD accretors. Table 1 lists all
cases
considered in the present work. 
case 1$^\star$ is considered as the standard model. In addition to
the nuclear-burning powered model, we consider an ``accretion
model'' that contains systems in which, under the assumptions of
case 1, the liberation of gravitational potential energy produces
symbiotic phenomenon ($L_{{\rm grav}}\geq 10L_\odot$) before the
first outburst of nuclear burning occurs or in the time intervals
between consecutive nuclear outburst plus decline
``quasi-cycles''.

\begin{figure*}
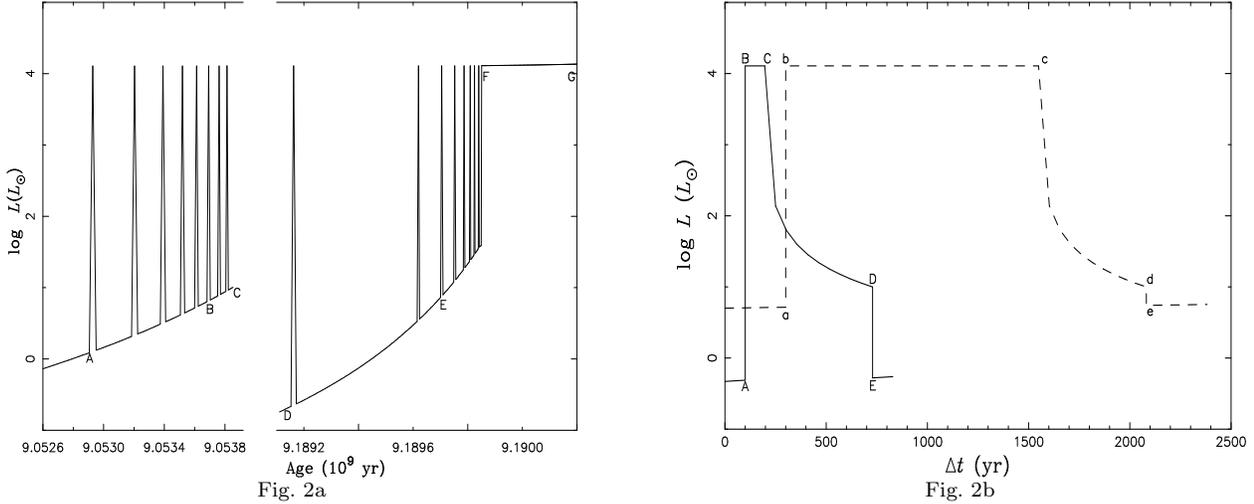

\begin{tabular}{c@{\hspace{3pc}}c}
\includegraphics[totalheight=3in,width=2.5in,angle=-90]{history.ps}&
\includegraphics[totalheight=3in,width=2.5in,angle=-90]{nova.ps}\\
Fig. \ref{history}a &Fig. \ref{history}b \\
\end{tabular}
\caption{---Luminosity of the primary component vs. evolutionary
           age in the system with initial masses of components 1.2 and 1.0
            $M_\odot$.
             Panel {\it a} represents the evolutionary
             ``history'' of the system as a SyS. Panel
             {\it b} shows the details of luminosity evolution during
             the first strong and weak novae experienced by the system  (solid and dashed
             lines, respectively). See text for details.
            }
\label{history}
\end{figure*}

\subsection{The history of a typical symbiotic binary}
\label{sec:history}

Our main results concerning the numbers of SySs and occurrence rates
of  SyNe are given in Table 2. But before we discuss them, we
present an evolutionary ``history'' of the typical binary that
becomes a SyS (Fig. \ref{history}). In this system the initial
masses of the primary and the secondary are 1.2 $M_\odot$ and 1.0
$M_\odot$ and their initial separation is 1000 $R_\odot$. We evolve
this binary under conditions appropriate to the ``standard'' case
1$^\star$. When the system attains point A in Fig.~\ref{history}a,
the primary has become a CO WD of $M= 0.54M_\odot$, the secondary is
in the FGB stage, the system still is not a SyS. At point A, the
first strong symbiotic nova occurs and the system becomes a SyS.
After the first strong symbiotic nova, the system leaves the state
of being a SyS, while accretion onto the WD continues until the next
symbiotic nova occurs. Before reaching point B, the system undergoes
5 strong SyNe. As the mass accretion rate increases, accretor enters
the regime of weak thermonuclear runaways at point B (in the latter
regime no erosion of WD is expected, see \S~\ref{maeff}). The system
experiences 3 weak SyNe, before the secondary leaves the FGB stage
at point C. After return to the giant branch but before reaching
point D, the secondary is an AGB star but the system does not
manifest itself as SySs. At point D, a strong symbiotic nova occurs.
After two strong SyNe, the system begins to undergo weak symbiotic
novae at point E. After 6 weak SyNe, the system enters the stage of
steady hydrogen-burning by the accretor at point F. At point G, the
secondary overflows its Roche lobe and the system leaves SySs'
stage. After the 4th outburst in the D-F stage in Fig~\ref{history},
in the inter-outburst time intervals, the star contributes to the
accretion model, since its $L$ never drops below 10\,$L_\odot$.

Figure~\ref{history}b shows in detail the evolution of the
luminosity during the  first strong and weak novae experienced by
the system. At point A (a), the outburst occurs and the luminosity
rapidly increases. From points B (b) to C (c), the system is in
the ``plateau'' stage. From points C (c) to D (d), the system is
in the decline stage. At point D(d) the luminosity of the accretor
drops below $10\,L_\odot$ and we no longer consider the system as
a SyS, since its accretion luminosity (represented by point E(e))
does not satisfy the criterion for SySs. Accretion continues until
critical ignition mass of hydrogen is accumulated again.
Figure~\ref{history}b also shows that the ``on''-stage for the weak
novae is much longer than that for the strong ones.

\subsection{ Birthrate and number of SySs }

\begin{table*}
\centering
\begin{minipage}{176mm}
  \caption{Different models of symbiotic stars population. The first column gives model number
           according to Table~\ref{tab:cases}.
           Columns 2 to 8 give the birthrates of SySs with accretors of different
           kinds (He-, CO- or ONe WD), the rate of SySs formation through  evolutionary
           channels I, II, and III (Fig.~\ref{channel}) and total birthrate.
           The ratio of the number of SySs in cooling
           stage and their total number is given in column 9. Columns 10 to 13 give the occurrence
           rates of SyNe with accretors of different kinds (He-, CO- or ONe-WD) and
           total rate, in the 13th column the numbers in parentheses mean the rates of weak
           SyNe. Total Galactic number of SySs with accretors of different
           kinds (He-, CO- or ONe-WD), the number of SySs formed through  evolutionary
           channels I, II, and III and total number are shown in
           columns 14 to 20, respectively.
           The 21st column gives the total numbers of Galactic SySs with
           cool components having apparent visual magnitude $\leq 12.0$. Columns 22 to 25 give the
           numbers for SyNe with accretors of different kinds (He-, CO- or
           ONe-WD) that are currently in the ``plateau'' stage of an outburst and their total number,
           in the 25th column the number in
           parentheses means the number of weak
           SyNe.}
\tabcolsep2.55mm
\begin{tabular*}{160mm}{|c|c|c|c|c|c|c|c|c|c|c|c|c|}
\cline{1-13}
\multicolumn{1}{|c|}{Model}&\multicolumn{7}{|c|}{Birthrate of
SySs(yr$^{-1}$) } & \multicolumn{1}{|c|}{N$_{\rm cooling}$} &
\multicolumn{4}{|c|}{Occurrence Rate of SyNe (yr$^{-1}$)}\\
\cline{2-8} \cline{10-13}
  &  He & CO & ONe &I & II & III& total&$\overline{\rm N_{total}}$&He&CO&ONe&total\\
\cline{1-13} 1&2&3&4&5&6&7&8&9&10&11&12&13\\
\cline{1-13} Standard&0.001&0.070&0.005&0.019&0.008&0.049&0.076&0.32& 0.002&1.4 &2.0& 3.4 (1.5)\\
Accretion&0.0&0.045&0.004&0.001&0.008&0.040&0.049&---& ---&--- &---& ---\\
case 2&0.004&0.080&0.006&0.033&0.008&0.049&0.090&0.38& 0.012&1.8 &2.3& 4.1 (1.7)\\
case 3&0.006&0.086&0.007&0.041&0.008&0.049&0.098&0.41& 0.022&2.1 &2.6& 4.7 (1.8)\\
case 4&0.018&0.106&0.006&0.074&0.008&0.049&0.131&0.33& 0.160&6.9 &6.4& 13.5 (6.0)\\
case 5&0.014&0.099&0.007&0.062&0.008&0.049&0.119&0.34& 0.120&3.9 &5.1& 9.2 (3.7) \\
case 6&0.0001&0.032&0.003&0.007&0.005&0.023&0.035&0.46&0.0001 &1.1&0.6& 1.7 (0.9)\\
case 7&0.001&0.121&0.007&0.019&0.008&0.102&0.129&0.62&0.002 &5.5&6.5& 11.9 (2.8)\\
case 8&0.001&0.078&0.006&0.014&0.004&0.067&0.085&0.59&0.002 &1.8&4.1& 5.9 (2.4)\\
case 9&0.003&0.074&0.005&0.026&0.006&0.050&0.082&0.26&0.007 &1.0&1.8& 2.8 (1.3)\\
case 10&0.00001&0.061&0.005&0.003&0.003&0.060&0.066&0.45&0.00001 &0.9&2.5& 3.4 (1.9)\\
case 11&0.0002&0.047&0.005&0.010&0.008&0.034&0.052&0.17&0.0002 &0.4&0.8& 1.3 (0.5)\\
case 12&0.004&0.089&0.006&0.027&0.008&0.064&0.099&0.52&0.011 &4.3&4.5& 8.8 (4.0)\\
case 13&0.001&0.071&0.005&0.016&0.008&0.053&0.077&0.46&0.001 &4.8&4.6& 9.5 (5.7)\\
case 14&0.001&0.070&0.005&0.019&0.008&0.049&0.076&0.37&0.002 &1.4&1.8& 3.2 (1.4)\\
\cline{1-13}
\end{tabular*}\\

---{\bf continue}\\

\tabcolsep2.30mm
\begin{tabular*}{160mm}{|c|c|c|c|c|c|c|c|c|c|c|c|c|}
\cline{1-13}
\multicolumn{1}{|c|}{Model}&\multicolumn{7}{|c|}{Number of
SySs } & \multicolumn{1}{|c|}{Number of }&
\multicolumn{4}{|c|}{Number of SyNe}\\
\cline{2-8}\cline{10-13}
 &  He & CO & ONe &I & II & III& total&
 SySs ($V_{\rm c}\leq 12.0$) & He&CO&ONe&total\\
\cline{1-13}1&14&15&16&17&18&19&20&21&22&23&24&25\\
\cline{1-13}Standard&{\bf $\leq$1}&3860&420&250&2280&1750&4300&19&{\bf $\leq$ 1}&407& 5&412 (393)\\
Accretion&0&6870&170&230&3730&3090&7100&37& ---&--- &---& ---\\
case 2&{\bf 10}&4220&570&780&2280&1750&4800&23&3&471& 5&480 (450)\\
case 3&{\bf 20}&4610&690&1290&2280&1750&5300&32& 10&563& 6&580 (543)\\
case 4&2060&11910&1080&11070&2280&1750&15100&73&599&1306& 15&1920 (1858)\\
case 5&820&9260&1120&7180&2280&1750&11200&59&304&1114&13&1432(1380)\\
case 6&{\bf $\ll$1}&1130&90&50&780&380&1200&4&{\bf$\ll$ 1}&103& 1&105 (99)\\
case 7&{\bf $\leq$1}&8830&2470&190&1680&9430&11300&74&{\bf $\ll$ 1}&1242& 9&1251 (1167)\\
case 8&{\bf $\leq$1}&2770&920&140&700&2860&3700&23&{\bf$\ll$ 1}&301& 7&309 (280)\\
case 9&{\bf $\leq$4}&3280&320&230&2120&1260&3600&12&{\bf $\leq$ 1}&197& 4&201 (182)\\
case 10&{\bf $\ll$1}&1600&360&30&340&1580&2000&6&{\bf$\ll$ 1}&146& 5&151 (140)\\
case 11&{\bf $\ll$1}&3220&240&100&1960&1420&3500&13&{\bf$\ll$ 1}&393& 5&398 (383)\\
case 12&{\bf 5}&5360&680&600&2970&2470&6000&32&{\bf$\leq$ 1}&411& 4&416 (395)\\
case 13&{\bf $\leq$1}&4980&420&250&2960&2180&5400&26&{\bf$\ll$ 1}&409& 4&413 (394)\\
case 14&{\bf $\leq$1}&3290&400&250&2030&1400&3700&21&{\bf$\ll$ 1}&406& 5&411 (392)\\
\cline{1-13}
\end{tabular*}
\label{tab:results}
\end{minipage}
\end{table*}

 First, we discuss the gross properties of the modeled population of
SyS and then proceed to the more detailed comparison of the
influence of different assumptions. As Table \ref{tab:results}
shows, the Galactic birthrate of SySs may range from about 0.035
(case 6) to 0.131 (case 4) yr$^{-1}$ and it is about 0.076
yr$^{-1}$ in the standard model. Their number is from about 1,200
(case 6) to 15,100 (case 4) and it is close to 4,300 in the
standard model. The contribution of SySs with He WD accretors to
the total number of SySs is negligible, due to their large
hydrogen ignition mass, except for cases 4 and 5 in which their
contribution is respectively close to 14\% and 7\%; the
contribution of SySs with ONe WD accretors strongly depends on
assumptions and the ranges from about 7\% (case 4) to 25\% (case
8); SySs with CO WD accretors contribute the main fraction  of the
population. Observational estimates of the total number of SySs
range from several thousands \citep{b2} to about 30,000
\citep{b21} or even up to 300,000 \citep{b31}, depending on the
assumptions on the distance to a typical SyS and on observational
selection.

 In the catalogue of \citet{b61}, there are
52 SySs with $V_{\rm c}\leq 12.0$ mag. In some cases $V_{\rm c}$
is the magnitude in outburst.
 In our models, the number of SySs with a cool component  brighter than
 $V_{\rm c}=12.0$ is from 4 (case 6) to 74 (case 7). In the standard model,
in combined nuclear and accretion models, there are in total 56
SySs within this magnitude limit, i.e., about 0.5\%  of all
systems! The situation is similar for simulations with other
assumptions. Thus, we may infer that a sample of observed
symbiotic stars that may be considered as statistically complete,
contains only fractions of a percent of all Galactic SySs.

In the nuclear models, from about 38 (case 7) to  83 (case 11)
percent of SySs are in the ``on''-state (including stably-burning
stars and stars in outbursts) and the rest are in the cooling
stage. When an outburst occurs, the hot component spends some time
in the ``plateau'' stage.  We assume that the time-span of the
``plateau'' stage is the lifetime of a SyNe. In total, the
fraction of the SyNe currently in the ``on''-stage among all SyS
(in the outburst and decline stages) varies from about 5\% (case
9) to 13{\%} (case 7).

 The contribution of
SyNe with He WD accretors to the total number of SyNe is about
31\% in case 4 and 21\% in case 5 but it is negligible in other
cases; due to very short lifetime of outbursts in SyNe with ONe WD
accretors, their contribution is also negligible (its range is
from approximately 0.4\% in case 7 to 2\% in case 8); most SyNe
have CO WD accretors. In the Galaxy, the model occurrence rate of
SyNe is between ranges from  about 1.3 yr$^{-1}$ (case 11) to 13.5
yr$^{-1}$ (case 4). The contribution of SyNe with ONe WD accretors
to the total rate is close to  60\% in most cases, in extreme
cases it is 74\% (case 10) and 35\% (case 6). The contribution  of
SyNe with He WD accretors to the total rate of Novae is
negligible. Note, the role of systems with ONe accretors is rather
uncertain because of uncertainty of their progenitors' range and
may be one of the reasons for the relatively high total occurrence
rate of SyNe in our models.

\subsection{Parameters}

{\bf 1. Common Envelope Evolution:} $\alpha$-algorithm --- In the
cases 2 and 3, the parameter $\alpha_{{\rm ce}}\lambda$ is
increased from 0.5 to 1.5 and 2.5, respectively. The larger
$\alpha_{{\rm ce}}\lambda $ is, the more easily the common
envelope is ejected, then the orbital period of a binary system
after common envelope phase is longer, which is favourable for the
formation of SySs. The occurrence rate of SyNe and the total
number of SySs in cases 2 and 3 become larger than  in the
standard model. On $\alpha_{{\rm ce}}\lambda$\ depends only
formation channel I (Fig.~\ref{channel}) in which the common
envelope is encountered. The number of systems produced through
this channel increases with $\alpha_{{\rm ce}}\lambda$\ since more
systems avoid merger in common envelopes. The number of SySs and
the occurrence rate of SyNe change with increased
$\alpha_{ce}\lambda$ by the same proportion as the birthrate in
channel I, since the outbursts are associated predominantly with
relatively more massive dwarfs which are produced mainly through
channels II and III.

{\bf 2. Common Envelope Evolution:} $\gamma$-algorithm --- In case
5, $\gamma$ is increased  to 1.75 from 1.5 in case 4. Based on Eq.
(5), the larger $\gamma $ is, the smaller the separation of the
binary is after the common envelope phase $a_{\rm f}$ (since more
orbital angular momentum is taken from the system). The birthrate
of SySs having undergone channel I in case 5 is lower than that in
case 4, but $\gamma$'s effect is weak.

However, there is great difference between $\alpha$-algorithm
cases (1, 2 and 3) and $\gamma$-algorithm cases (4 and 5). The
birthrate of SySs formed through channel I in case 4 is about 3.9
times of that in the standard model, the birthrate of SySs with He
WD accretors in case 4 is about 18 times of that in the standard
model, and the total birthrate of SySs in case 4 is 1.7 times
higher than that in the standard model. The occurrence rate of
SyNe and the total number of SySs in case 4 are much larger than
those in the standard model. The reason is that with
$\gamma$-algorithm post-common-envelope systems are wider than
with $\alpha$-algorithm and this facilitates symbiotic phenomenon,
allowing more stars to evolve further along FGB or AGB before the
second RLOF. The $\gamma$-mechanism is especially favourable for a
larger fraction of SySs with He-WD accretors among all SyS since
it (i) allows the progenitors of He-WDs to avoid merger, (ii)
hydrogen shells accreted by He-WDs are more massive, while their
luminosity is lower, hence, they live as SySs longer. Also, in
case 4 more massive white dwarfs survive common envelopes and this
results in a sharp increase in the occurrence rate of SyNe.

3. {\bf Parameter $v_\infty$} --- The comparison of cases 1, 6, 7
and 8 shows that $v_\infty$ is a key factor for forming SySs. For
a given mass loss rate, the decrease of $v_\infty$ increases the
efficiency of accretion, facilitating SyS formation. This is best
exemplified by comparison of cases 1 and 6. In case 7 in early AGB
stage, the accretion efficiency of a WD is enhanced due to low
$v_\infty$, but the mass loss rate of a cool giant is not high. In
late AGB stage, the mass loss rate of a cool giant is high, but
the accretion efficiency of a WD is low due to rapidly rising
$v_\infty$. As a result of these effects, the rate of accretion
onto WDs stays in the range which is favourable for the occurrence
of SyNe (see Fig.~\ref{MA} and Table 2).

4. {\bf Parameter $\alpha_{\rm w}$}: --- In case 9, the wind
velocity law suggested by \citet{b52} is adopted. This empirical
law gives very low velocities in the vicinity of the cool
component, which results in more efficient accretion and hence
increases birthrate and number for the closest systems with a
He-WD, but slightly reduces them for systems with CO- and ONe-WDs.
However, for $\alpha_{\rm w}=1$ in case 10 wind velocity in the
vicinity of the accretor is high. This strongly decreases the
birthrates of SySs formed through  channels I and II. Also, SySs
with He-WD accretors are hardly formed. For the systems formed
through channel III the birthrate and number almost do not depend
on $\alpha_{\rm w}$, since for wide systems $\alpha_{\rm w}$ is
close to 1 for all wind-acceleration laws.

5. {\bf Parameter $\Delta M_{\rm {WD}}^{{\rm crit}} $} --- In
cases 12 and  11, $\Delta M_{\rm {WD}}^{{\rm crit}}$, the
threshold value for ``ignition'' of symbiotic novae, are,
respectively,  $\frac{1}{3}$ and 3 times of that in the  standard
model. The occurrence rate of SyNe and total number of SySs
rapidly increase from case 11 to the standard model and to case
12, but the numbers of SyNe and SySs in the burning stage are
basically invariable, since the amount of hydrogen available for
burning is the same in  all cases. In case 13, Eq. (A1) of
\cite{b38} is used for $\Delta M_{\rm {WD}}^{{\rm crit}} $.
Compared with Eq. (\ref{eq:yunm}), it agrees within a factor close
to 2 when the mass accretion rate is $10^{-9}M_\odot$yr$^{-1}$.
However, Eq. (A1) of \citet{b38} gives $\Delta M_{\rm {WD}}^{{\rm
crit}}$ that are, on the average, 3 times lower than the ones
given by Eq. (\ref{eq:yunm}) when the mass accretion rate is
$10^{-8}M_\odot$yr$^{-1}$ and 7 times lower when the mass
accretion rate is $10^{-7}M_\odot$yr$^{-1}$. So the occurrence
rate of SyNe increases.

6. {\bf Optically thick wind} --- Comparing cases 1 and 14, we
find that it is practically unimportant for the model of the
population of SySs whether there is an optically thick wind or
not. SyNe are not affected, since they occur at low $\dot{M}$. In
case 14, the total number of SySs decreases because in the absence
of optically thick winds expansion of accretors and formation of
common envelopes in the highest $\dot{M}$ systems becomes
unavoidable. The outcome of common envelope is either merger of
components or formation of a double-degenerate.

To summarize, we find that  the assumed formalism for the common
envelope evolution, terminal wind velocity and $\Delta M_{\rm
{WD}}^{{\rm crit}}$ have the strongest effect upon the occurrence
rate of SyNe and total number of SySs, introducing uncertainty up
to factors $\simeq (3 - 4)$. The wind velocity law affects  the
occurrence rate of SyNe and the number of SySs within factor $\leq
2$.

\subsection{Properties of SySs  }
In this section, we describe potentially observable physical
quantities of SySs. We compare the standard model
($\alpha$-algorithm for common envelope evolution), case 4
($\gamma$-algorithm for common envelope evolution), case 7 (low
outflow velocity for AGB stars), and the accretion model.

\begin{figure}
\begin{tabular}{c}
\includegraphics[totalheight=3.2in,width=3.5in,angle=-90]{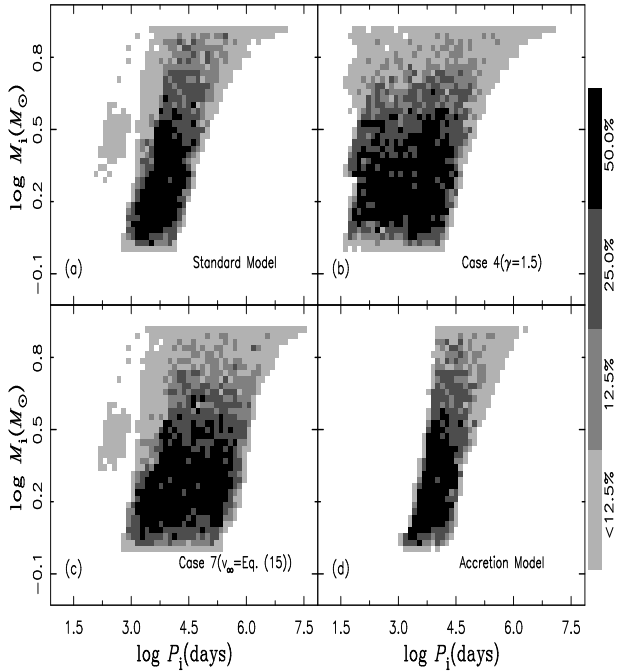}

\end{tabular}
\caption{---Gray-scale maps of initial primary
            mass $M_{\rm i}$ vs. initial orbital period $P_{\rm i}$ distribution
            for the progenitors of SySs. The gradations of gray-scale
            correspond to the regions where the number density of systems is,
            respectively,  within 1 -- 1/2,
            1/2 -- 1/4, 1/4 -- 1/8, 1/8 -- 0 of the maximum of
             ${{{\partial^2{N}}\over{\partial {\log P_i}}{\partial {\log
            M_i}}}}$, and blank regions do not contain any stars.
           The cases shown in particular panels are indicated in their lower right corners.
            }
\label{Initial}
\end{figure}

Figure~\ref{Initial} shows the distributions of the progenitors of
SySs,  in the ``initial primary mass -- initial orbital period''
plane. Most SySs descend from $ (1.0 - 2.5)\,M_\odot$ stars, this
is an effect of the IMF. The Figure also shows well that for
$\alpha$-algorithm for common envelopes, in most SySs, accretors
form from AGB stars. The distribution  $\log M_{\rm i} - \log
P_{\rm i}$ in case 4 ($\gamma$=1.5) is different from that in
other cases, because for the $\gamma$-algorithm, common envelopes
result in rather moderate shrinkage of orbits or even in their
expansion, allowing relatively close systems to avoid merger, as
already noted above. In this case also stars in FGB contribute to
the formation of SySs, particularly, by formation of He-WD
accretors and relatively low-mass CO-WD accretors, see also
Fig.~\ref{HOTM} below. The distribution becomes more even in case
7 because under assumption of low $v_\infty$ in the systems formed
through Channel III velocity of the matter passing by WD is lower
than in the ``standard'' case and accretion becomes more
efficient. For the ``accretion'' model the distribution has more
narrow range of periods since, in fact, the objects of this model,
are a subsample of standard model which had  $L>10\,L_\odot$\
prior to the first outburst or between outbursts.

\subsubsection{ Orbital Periods}

\begin{figure}
\begin{tabular}{c}
\includegraphics[totalheight=3.2in,width=3.5in,angle=-90]{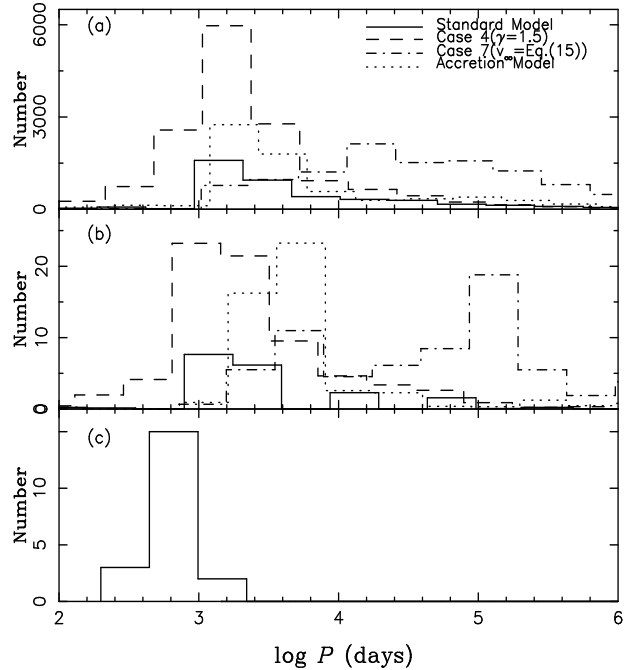}
\end{tabular}
\caption{---Number distributions of SySs as a function of
             orbital periods.
             Panel {\it a} shows  distributions in total samples for standard model,
             accretion model and cases 4, and 7, while panel {\it b} shows
             distribution for a sub-sample with cool components
             brighter than $V_{\rm c}=$12.0 mag. See the key to lines in panel a. In panel
             {\it c} observational data from \citet{b29} and \citet{b61} is plotted.}

\label{TB}
\end{figure}

Figure~\ref{TB} shows the distributions of SyS over orbital
periods. The plot for the total sample, panel {\it a}, shows that
in case 4 ($\gamma$-algorithm) the distribution is wider than in
the standard case ($\alpha$-algorithm) since in the former case
there are systems which avoided merger and retained or even
increased their initial separations. In case 7, there are more
SySs with long orbital periods than in the standard case thanks to
the systems that form through Channel III and manifest symbiotic
phenomenon because of the low wind velocity which enhances
accretion efficiency.

The distribution of observed SyS with $V_{\rm}\leq 12.0$ mag. over
orbital periods based on the data from \citet{b61} and \citet{b29}
is plotted in Fig.~\ref{TB}c. It may be compared with the
distributions for the model sub-samples of systems with
$V_{\rm}\leq 12.0$ mag. (Fig.~\ref{TB}b). The overwhelming
majority of case 1
 and case 4  model systems with
V$<12$ in Fig.~\ref{TB}b are almost uniformly distributed between
1000 and 6000.
 However,
orbital periods are measured for less than 30 systems out of
almost 200 known SySs and it is hard to measure long orbital
periods. The predictions of models about orbital period can be
hardly verified. However, Fig.~\ref{TB}
 suggests that
SySs with  orbital periods shorter than 200 day are hardly
expected, which is in agreement with observations.

\subsubsection{ Properties of  Hot Components}

\begin{figure}
\begin{tabular}{c}
\includegraphics[totalheight=3.2in,width=3.5in,angle=-90]{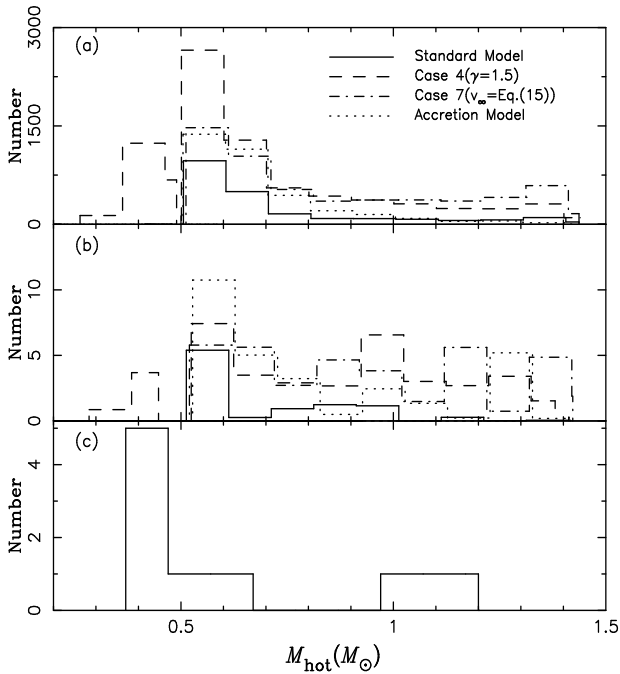}
\end{tabular}
\caption{---Number distributions of model Galactic SySs  as a
function of
            the hot component mass.
            Fig.~\ref{HOTM}a is for the total samples,
            Fig.~\ref{HOTM}b is for a sub-sample with cool components
            brighter than $V_{\rm c}=$12.0 mag, and
            Fig.~\ref{HOTM}c
            is plotted based on the data in  \citet{b29} and \citet{b61}.
            The key to the line-styles representing different models is given
            in the upper right corner of Fig.~\ref{HOTM}a.
}
\label{HOTM}
\end{figure}
\begin{figure}
\begin{tabular}{c}
\includegraphics[totalheight=3.2in,width=3.5in,angle=-90]{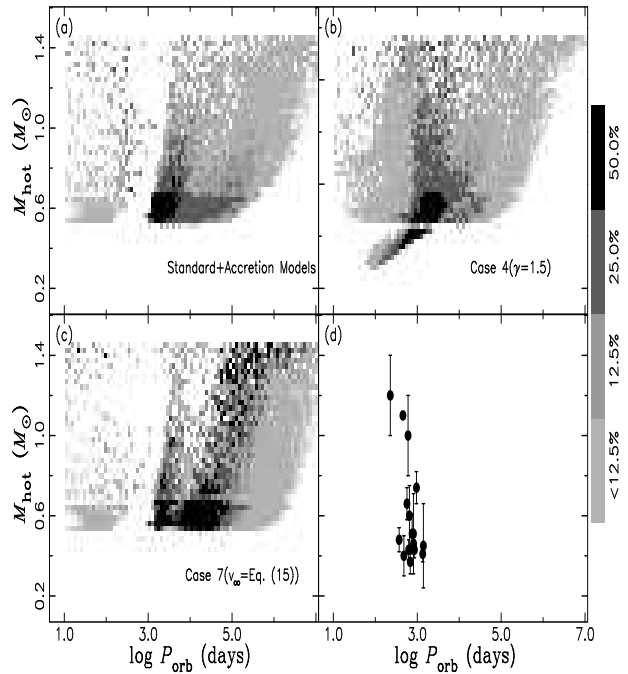}
\end{tabular}

\caption{---Gray-scale maps of  the model distributions  of the
           masses of hot components
            vs. orbital periods. The gradations of gray-scale
            correspond to the regions where the number density of systems is,
            respectively,  within 1 -- 1/2,
            1/2 -- 1/4, 1/4 -- 1/8, 1/8 -- 0 of the maximum of
             ${\partial^2{N}} \over {\partial {\log P_{\rm orb}}{\partial {\log
            M_{\rm hot}}}}$, and blank regions do not contain any stars. Fig.~\ref{graypm}d
            is based on the data for observed systems
            from \citet{b29}.
            }
            \label{graypm}
\end{figure}
\begin{figure}
\begin{tabular}{c}
\includegraphics[totalheight=3.2in,width=3.5in,angle=-90]{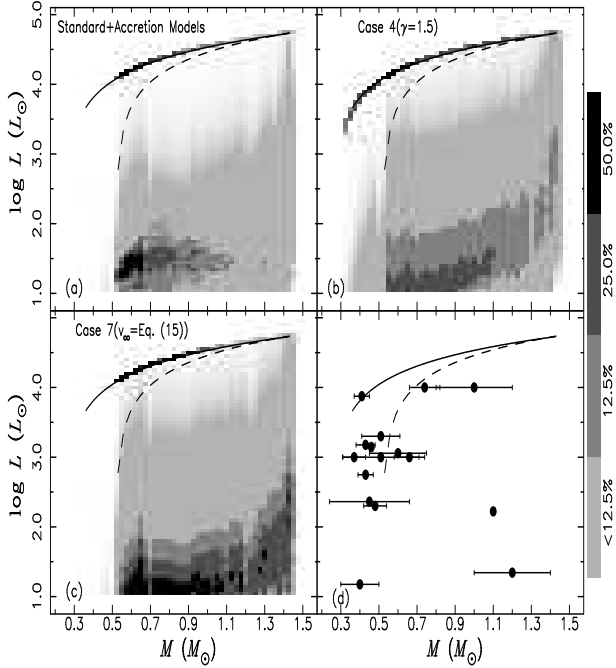}
\end{tabular}

\caption{---Gray-scale  maps of the distribution of luminosities
            of hot components in SySs
            vs. their masses. The gradations of gray-scale
            correspond to the regions where the number density of systems is,
            respectively,  within 1 -- 1/2,
            1/2 -- 1/4, 1/4 -- 1/8, 1/8 --0 of the maximum of
             ${{{\partial^2{N}}\over{\partial {\log L}}{\partial {\log
            M}}}}$, and blank regions do not contain any stars. Fig.~\ref{grayml}d shows $M_{\rm hot}$ estimates
            from \citet{b29}.
            Solid line shows relation between mass of accreting WD
            and its luminosity (\ref{eq:mcl}), dashed line is
            \citep{l11,l12} mass-luminosity relation for AGB stars
            $L/L_\odot=6.0\times10^4(M_{\rm c}/M_\odot-0.52)$.}
           \label{grayml}
\end{figure}
\begin{figure}
\includegraphics[totalheight=3.2in,width=3.5in,angle=-90]{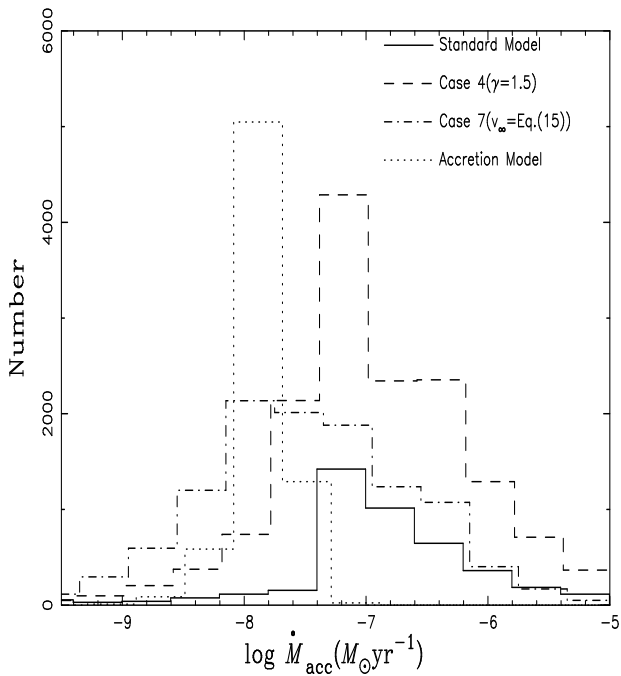}

\caption{---Number distributions of model SySs as a function of
            the mass-accretion rate of the hot components.
            }
\label{MA}
\end{figure}

In Fig.~\ref{HOTM}, the distributions of the masses of hot
components are shown. The distribution for case 4 ($\gamma$=1.5)
is different from other cases. In case 4, there is a  peak between
approximately 0.35 and 0.5 $M_\odot$ which represents the
distribution of masses of He WD accretors. Note, the masses of
these dwarfs are close to the upper limit of the masses of He WDs.
In other cases, the contribution of SySs with He WD accretors is
negligible. For all nuclear models, the distributions have   peaks
around $ 0.6\,M_\odot$ and a slight enhancement by 1.35 $M_\odot$.
 The
peak is due to the distribution of CO WD masses. The enhancement
at the tail of the distribution is due to the massive WD accretors
(especially ONe WDs) that undergo many outbursts because of low
hydrogen ignition mass. The model distribution for the sub-sample
for SySs with cool components brighter than 12.0 mag.
is shown in Fig.~\ref{HOTM}b. The distribution of the estimated
masses of hot components in systems with
$V_{\rm c} \leq 12.0$ mag
[as given in \citet{b61} and \citet{b29}],
 is plotted in Fig.~\ref{HOTM}c.
Though it is still small number statistics, we should note that
only with the $\gamma$-formalism for common envelopes is it
possible to obtain the prominent fraction of low-mass accretors
which is present in the observed population. Note the presence of
relatively massive WD-accretors also in the observed population,
though not so prominent as in case 7 and not extending to the
range of masses of ONe WDs. The excess of massive WDs, if real,
may be a counterargument to the case 4 model.

Figure \ref{graypm}  presents, in gray scale, the relation between
the masses of the hot components  and orbital periods of SyS. The
distributions in Figs.~\ref{graypm}a and c, are split into two
domains. The scarcely populated left one represents SySs formed
through channel I; the right domain is populated by systems formed
through channels II and III. The gap between the two domains is
absent in case 4 in which systems that went through channel I did
not experience a sharp decrease of orbital separation. Panel c
shows that a low velocity stellar wind favours symbiotic
phenomenon in systems within a wider range of orbital separations
than in the models with high velocity.

The comparison with observed systems shows that our models correctly
predict that the majority of hot components of SySs must have masses
around (0.5-0.6)\,$M_\odot$, though the existence of some lower mass
(He-WD) accretors cannot be excluded. Note that mass estimates
typically assume orbital inclinations of $i=90^\circ$  or a limit to
$i$ (see Table 2 in \citet{b29}), hence, the estimates are lower
limits. Orbital periods of the systems with measured $M_{hot}$ are
also consistent with theoretical expectations. At the same time our
models suggest the existence of a significant number of SyS with yet
unmeasured periods larger than $ 1000$ day.

\begin{table}
   \caption[]{An example of an evolutionary scenario
leading to
   formation of T CrB-like system. Masses are in
$M_\odot$, $P_{\rm
   orb}$ in days. The values of binary parameters at the
onset of respective
   stage are given. } \tabcolsep1.mm
   \begin{tabular}{ccccccc}
    \hline
    \hline
& MS,MS & AGB,MS  & CE & WD,MS &WD,FGB & CE  \\
    \hline
    \hline
    $M_1$ & 7.09      & 6.83         & 2.19  &1.28       &
1.28        &1.28\\
    $M_2$ & 0.90      & 0.90        &1.25    & 1.25     &
 1.25      &1.19\\
    $P_{\rm orb}$ &7817    &8155     &10622   &277    &277
  & 263  \\
    \hline
    \label{tab:tcrb}
    \end{tabular}
   \end{table}

Panel \ref{graypm}d shows that there is a group of
 ``outliers'' -- systems with relatively short
periods and massive  accretors. Among them is the prominent
recurrent nova system T~CrB ($P  \approx 227.6$ day, $M_{\rm
hot}=1.2\pm0.2\, M_\odot$, $M_{\rm cool}=0.7\pm0.2\, M_\odot$).
Progenitors of such systems must have high initial mass  primaries
and be sufficiently wide to form a massive WD and have large
initial mass-ratios of components to enable sufficient  shrinkage
of the orbit in common-envelope event. With the notation  already
used above, we present a numerical example for formation and
evolution of  a system that may belong to this group (in the
standard model) in Table~\ref{tab:tcrb}. The system  manifests
symbiotic phenomenon in the ``WD,FGB''-stage. The giant  loses
only  about $ 0.06M_\odot$\ of the matter before RLOF and  the WD
experiences about $ 1500$\ strong Novae (this number is rather
uncertain because of the uncertainty in $\Delta M^{\rm WD}_{\rm
crit}$). Note, the mass of WD practically does not change.

Figure \ref{grayml} shows the relations between  orbital periods of
SyS and  the luminosities of their hot components. By our definition
of SySs, there are 3 states of WDs in them, see \S~\ref{sec:history}
and Fig.~\ref{history}. Steady-burning systems and SyNe systems in
the plateau stage obey the mass-luminosity relation~(\ref{eq:mcl}).
These systems are located in Fig.~\ref{grayml} along the solid line
representing Eq.~(\ref{eq:mcl}). Then, when they are in the decline
stage, they deviate from (\ref{eq:mcl}) downward. Stars in the
plateau and decline stages form the upper populated area in
Fig.~\ref{grayml}a,b,c. Next, as Fig.~\ref{history} shows, stars
rapidly transit into an ``accretion'' stage where they spend time
comparable or even longer than in the ``nuclear-burning'' stages.
This explains the formation of the lower populated domain in
Fig.~\ref{grayml} and the existence of a gap between the two
regions. The lower cut-off at $L/L_\odot=10$ corresponds to our
definition of SySs as systems with a larger luminosity WD. The
difference in the density of systems in different regions of
Fig.~\ref{grayml}a,b,c may be understood as a consequence of
different period distributions and accretion efficiencies in
different cases. This picture is  reasonably consistent with
observations plotted in Fig.~\ref{grayml}d which shows that most
observed systems are in the decline stage, but, probably,
observational selection favours detection of  the brightest systems
only\footnote{\citet{b16} also suggest that observed symbiotic stars
are in the stage when their luminosity is provided by residual
nuclear burning and cooling luminosity of WD.}. Stars in the long
accretion stage are not observed because of their low luminosities.
Because hot components of the observed systems in the
mass-luminosity plot are located between Iben-Tutukov and
Paczy\'{n}ski-Uus curves but not below the latter curve,
Fig.~\ref{grayml}d also suggests that mass-luminosity relation for
{\it hydrogen-accreting} WDs is really close to Iben-Tutukov
relation~(\ref{eq:mcl}) and differs from Paczy\'{n}ski-Uus relation
$L/L_\odot=6\times10^4(M_c/M_\odot-0.52)$ which describes the
luminosity of {\it AGB stars with double nuclear-burning shells.}

In Fig.~\ref{MA}, the distributions of mass accretion rates onto the hot
components $\dot{M}_{{\rm acc}}$ are shown. For all nuclear models the
distributions have peaks close to $ 10^{-7}M_\odot$ yr$^{-1}$;  for the
accretion model the peak is close to  $10^{-8}M_\odot$ yr$^{-1}$ . The first
peak and its extension to the larger $L$ correspond to the rate of stable
hydrogen burning at the surface of typical WD accretors, while the second peak
corresponds to the release of 10$L_\odot$ by accretion.

During an outburst, the hot component may support an additional high-velocity
wind [e.g., \citet{b19,b16}]. As noted by the latter authors, wind mass-loss
may shorten the lifetime of SyS in the ``on''-state. Another related effect is
the formation of a temporary ``common envelope'' by an extended envelope of the
white dwarf during outburst and associated mass loss. We neglect here these two
effects. Note, collision of the winds from components is important for the
formation of observational features of SySs, but it does not influence their
evolution.

\subsubsection{ Properties of  Cool Components}
\begin{figure}
\begin{tabular}{c}
\includegraphics[totalheight=3.2in,width=3.5in,angle=-90]{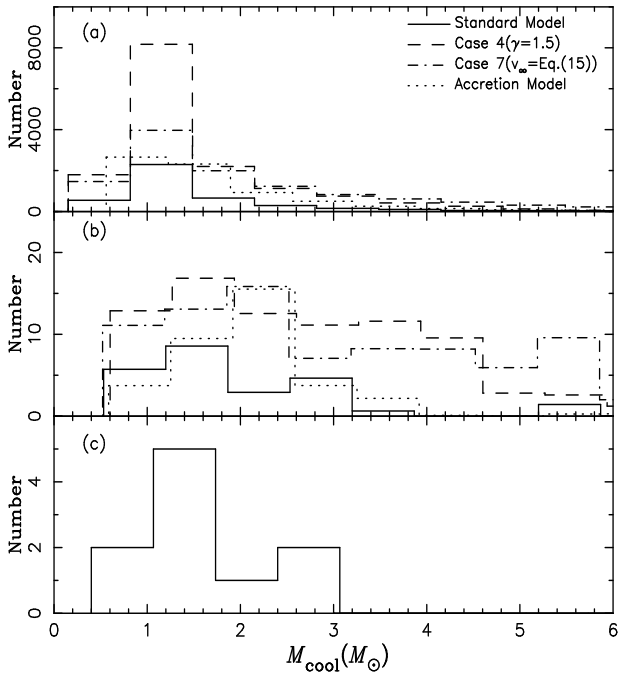}
\end{tabular}

\caption{---Number distributions of model SySs
            as a function of
            the mass of the cool components.
            Panel a is for the total samples, panel b
            is for a sub-samples with cool components
            brighter than $V_{\rm c}=$12.0 mag and
            panel c presents data from \citet{b61} and \citet{b29}.}
\label{COOLM}
\end{figure}
\begin{figure}
\begin{tabular}{c}
\includegraphics[totalheight=3.2in,width=3.5in,angle=-90]{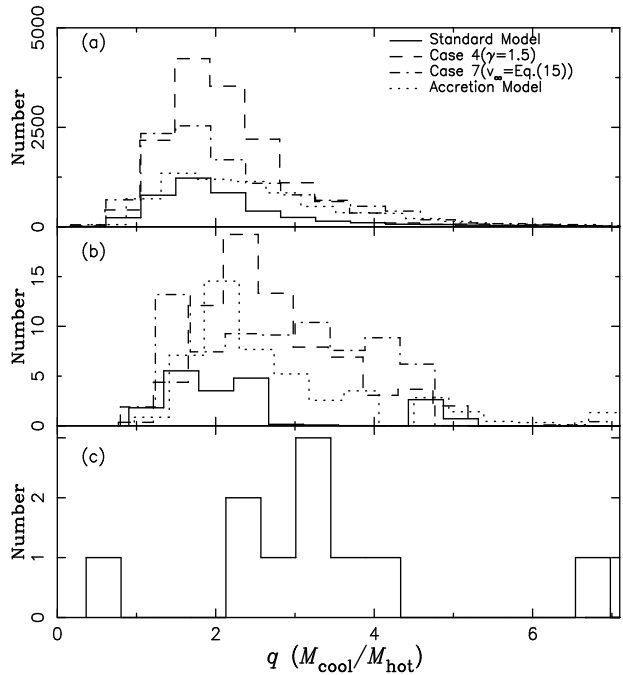}
\end{tabular}
\caption{---Number distributions of model SySs as a function of
            the mass ratio of components.
            Fig.~\ref{Q}a is for the total samples, Fig.~\ref{Q}b is for a sub-sample with cool components
            brighter than $V_{\rm c}=$12.0 mag., and
            Fig.~\ref{Q}c presents estimates of $q$ in observed systems after \citet{b61}
            and \citet{b29}.}
\label{Q}
\end{figure}
\begin{figure}
\begin{tabular}{c}
\includegraphics[totalheight=3.2in,width=3.5in,angle=-90]{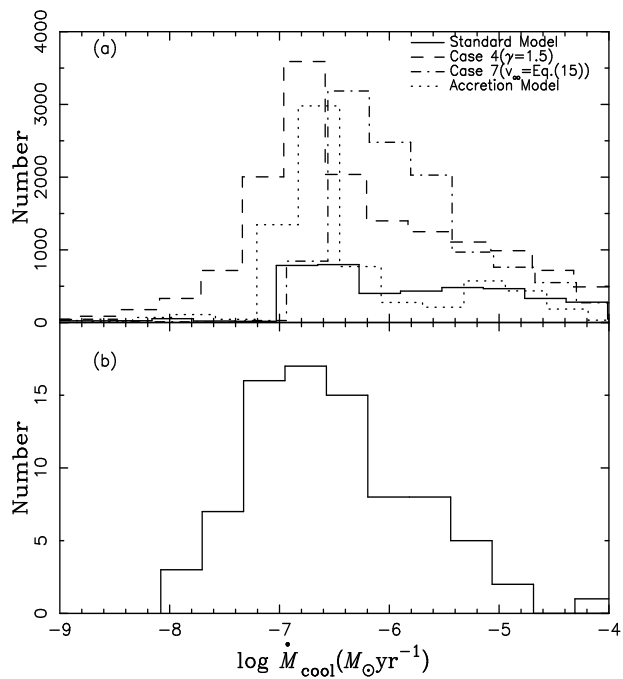}
\end{tabular}

\caption{---Number distributions of model SySs
            over mass-loss rate of the cool component.
            Fig.~\ref{ML}a is for total model samples.  Fig.~\ref{ML}b
            presents the observational data from \citet{b46}.}
\label{ML}
\end{figure}
\begin{figure}
\includegraphics[totalheight=3.2in,width=2.5in,angle=-90]{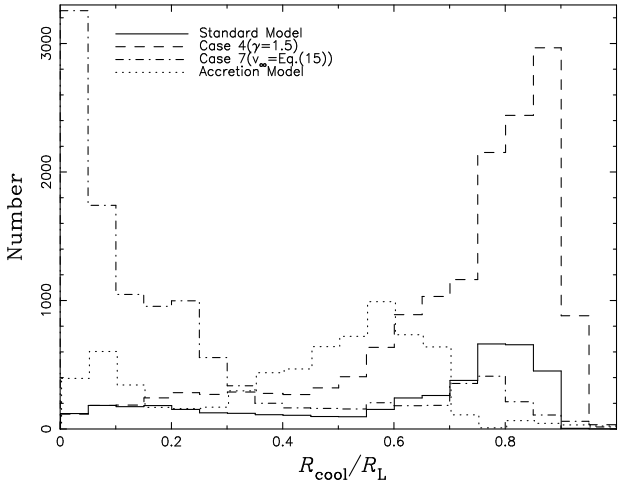}
\caption{---Number distributions of model SySs
            over the ratios
            of donor radius to the Roche lobe radius of cool component.
            }
\label{RL}
\end{figure}

The cool component supplies the matter accreted by the hot one. The mass of the
cool component gradually decreases. In Fig.~\ref{COOLM}a, the distributions of
cool components' masses are shown. In all cases, the great majority of masses
are confined to the range   $(0.8 - 2.0)\,M_\odot$. The distributions for the
model sub-samples  with cool components brighter than $V_{\rm c}=$12.0 mag are
shown in Fig.~\ref{COOLM}b, the distribution of the masses of observed cool
components of SySs with $V_{\rm c}\leq 12.0$ mag.  \citep{b61,b29} is plotted
in Fig.~\ref{COOLM}c.  Results of our  simulations  reasonably agree with
observations.

Figure~\ref{Q}a shows the distributions of mass ratios of cool components to
hot components for the total model samples. The distributions for the model
sub-samples with cool components brighter than $V_{\rm c}=$12.0 mag. are shown
in Fig.~\ref{Q}b.  The peaks in Fig.~\ref{Q}a for all cases are close to 1.8.
Figure~\ref{Q}c shows the distribution of the observational estimates of  $q$
for  systems with $V_{\rm c}\leq 12.0$ mag. from \citet{b61,b29}. The peak in
Fig.~\ref{Q}b is located between 2 and 4. We do not consider this discrepancy
as critical. First, it is based on very small number statistics. Second, the
masses for observed cool components are typically assigned based on their
spectral types and evolutionary tracks, but for M-giants the
masses within the same spectral subtype may differ by a factor larger
than 2 \citep{ds98}.

In Fig.~\ref{ML}, the distributions of SySs over mass-loss rates from the cool
components are shown. The peaks in Fig.~\ref{ML}a for the total samples are
between $ 10^{-7}$ and $10^{-6}M_\odot$yr$^{-1}$. The observational estimates
of the mass-loss rates after \citet{b46} are plotted in Fig.~\ref{ML}b.
Comparison of Figs.~\ref{ML}a and~\ref{ML}b shows that  our results agree
reasonably well with observations.

Figure~\ref{RL} shows the distributions of the cool components
over the ratio of their radii to the radii of their Roche lobes.
In Fig~\ref{RL}, the distribution for case 7 is different from
other nuclear cases. In case 7, the maximum is below 0.3, but it
is between 0.7 and 0.9 in the standard model and case 4. The main
reason for the difference is velocity of the stellar wind. In the
standard model the rate of mass-loss increases with evolutionary
lifetime, $\alpha_w$ evaluated by Eq.~(\ref{eq:windyl}) decreases,
thus,  the model favours symbiotic stars with donors close to the
Roche lobe. In case 7, velocity of the stellar wind is low during
the early AGB stage and this favours symbiotic stars with
relatively compact donors.

\citet{b34} found that M-type giants in symbiotic binaries obey the relation
$R\leq l_1/2$ where $R$ is the radius of the cool component and $l_1$ is the
distance from the center of the cool component to the inner Lagrangian point
$L_1$. Note, it is difficult to compare  Fig.~\ref{RL} directly with the
observational data because Sp-M, Sp-R relations are not unique for stars in the
late stages of their evolution. \citet{b34} applied, for the radii of giants,
the median values from \citet{ds98}, although the real radius may be several
times different. However, if the pattern found by \citet{b34} is really
correct, the model of SySs will require a rather low velocity of stellar wind
in the low mass loss rate phase but a high wind velocity in the high mass loss
rate phase, which is consistent with the stellar wind model of \citet{b53}.

\subsubsection{Symbiotic Novae }

\begin{figure}
\begin{tabular}{c}
\includegraphics[totalheight=3.0in,width=3.5in,angle=-90]{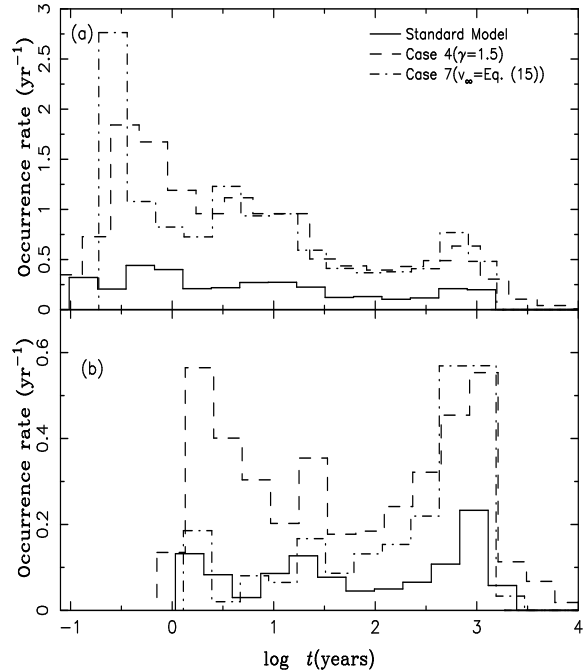}
\end{tabular}
\caption{------Occurrence rate of SyNe (Fig.~\ref{NOVA}a) and
               weak SyNe (Fig.~\ref{NOVA}b) as a function of the
              duration  of outbursts.
 }
 \label{NOVA}
\end{figure}

For the nuclear models, the most important property of SySs is
their thermonuclear runaways. The length of the outbursts depends on
the
mass and
mass-accretion rate of the WD-component.
As Fig.~\ref{NOVA}a shows, SyNe last from several months to
thousands of years. The distributions in Fig.~\ref{NOVA}a can be
roughly separated into two domains.
In the first domain, below 1 yr,  there are peaks around  4
months. The first domain corresponds to the strong SyNe. The
second domain extends  from about 1 yr to several
hundred (even several thousand) yr. As Fig.~\ref{NOVA}b shows, to
this domain contribute, predominantly, weak  SyNe.

In the standard model for the population of Galactic SyNe, the
contribution of  weak SyNe to the  total occurrence rate of SyNe
is approximately 40\%; it ranges from 24\% in case 7
[$v_\infty$=Eq. (\ref{eq:winters})] to 56\% in case 10
($\alpha_{\rm w}=1$). If we consider the number of systems
experiencing thermonuclear runaways, it appears  that, due to the
shorter lifetime of strong SyNe, the contribution of systems with
weak SyNe to the total number of SyNe-systems is larger than 90\%.

Figure~\ref{NOVA} shows that some outbursts last for more than
1000 years. These outbursts occur on low-mass He-WDs. This is
especially prominent in case 4 ($\gamma$=1.5) in which a large
number of He WD accretors form. However, these long outbursts are
more difficult to detect.  Observed outbursts last for months to
years or for dozens of years \citep{b19,b29}. The outburst of AG
Peg lasted for 100 years. Thus the double-peak structure of the
distribution in Fig.~\ref{NOVA}a is, crudely, consistent with
observations. We suggest the existence of two populations of SySs
-- those with short outbursts and those with long ones.

Note that the observed flashes of SyNe with small amplitude (1 to
3 mag) may also be  due to the variations  of mass-transfer rate
and/or accretion disk instabilities and associated variations in
the nuclear burning rate \citep{d86,l18,l13,s06}.

For the occurrence rate of SyNe, \citet{b16}, taking into account
the fact that four outbursts were observed within 4 kpc from the
Sun in the past 25 years, obtained a mean value of $\nu \sim 3$
yr$^{-1}$. From the number of SyNe's ``fossils'', \citet{l15}
estimated the frequency of SyNe as about 0.4 yr$^{-1}$, which may
be considered as the lower limit. Our model estimates are between
$ 1.3$ yr$^{-1}$ and $13.5$ yr$^{-1}$. In the standard model, the
occurrence rate of SyNe is close to 3.4 yr$^{-1}$, which agrees
with \citet{b16} estimates. Comparing our results to the
observational estimates, one should note the following: first, not
all SyNe are observed; second, it is difficult to confirm the
event of SyNe if it lasts for more than 100 yr. The problem of the
incompleteness of the sample of Symbiotic Novae was, in fact,
never explored.

\subsubsection{Symbiotic Stars and Supernovae Ia}

All models for Supernovae Ia (SNe Ia) involve accreting WDs. The
 efficiency of the accumulation of helium shell mass is very
important if we consider SySs as the potential progenitors of SNe
Ia. Helium flashes may occur in our models because the critical
ignition helium mass is very low (See Eq. (\ref{eq:Hemc})) when
the mass accumulation rate of helium is higher than
$10^{-8}M_\odot$yr$^{-1}$. The range of their occurrence rate is
between about 0.0006 yr$^{-1}$ (case 6) and 0.008yr$^{-1}$ (case
4), and the occurrence rate in the standard model  is close to
$0.005$yr$^{-1}$.

\begin{figure}
\includegraphics[totalheight=3.0in,width=2.5in,angle=-90]{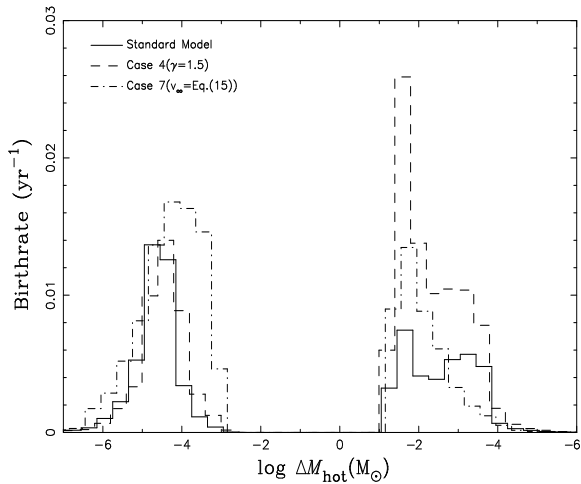}
 \caption{---Occurence rate distributions of SySs as a function of the 
mass
             changes of CO WD accretors. The bins to the left of 0 at x-axis show 
             eroding WD, while to the right of 0 -- mass-accumulating WD.
  }
 \label{BURNH}
\end{figure}

\citet{b20}, \citet{b60}, \cite{b166} and \cite{b266}
showed that it is unlikely that SySs can produce SN Ia via
accumulation of the Chandrasekhar mass.
Our results agree with this conclusion. In all cases, the
frequency of events in which a CO WD reached the Chandrasekhar
limit is lower than 10$^{-6}$yr$^{-1}$.
The reason is the
lack of massive accreting CO WDs and the inefficient increase of
core masses (Fig.~\ref{BURNH}) and even erosion of accretors
(especially, for  massive WD).
This is clearly seen from Fig.~\ref{BURNH}:
accumulated helium shell
masses do not exceed 0.1$M_\odot$.

The low efficiency of helium accumulation also makes symbiotic
stars bad candidates for SNe Ia produced by ``edge-lit detonation
(ELD)'' --- detonation of core carbon initiated by detonation of
accreted helium in WDs in the mass range 0.6---0.9$M_\odot$, after
accreting 0.15---0.2$M_\odot$ of He at $\dot{M} \sim 10^{-8}
M_\odot {\rm yr}^{-1}$. [e.g., \citep{b26,b24,b56}]. In all our
models, the detonation of accreted helium can not occur, that is,
there is not a sample in which accumulated helium shell mass
exceeds 0.15$M_\odot$. We calculated the rate of events in which a
$0.1M_\odot$ helium shell is accumulated on CO WDs with masses
larger than 0.6$M_\odot$ in all models. This rate is $ < 10^{-5}$
yr$^{-1}$. Furthermore, as we mentioned above, in the case of fast
rotation, outbursts of accreted He may occur after accretion of
only several 0.01 $M_\odot$ of matter \citep{b82}. Then, ``Helium
Novae'' may occur.

\subsection{Comparison with YLTK}

YLTK performed a detailed population-synthesis study of SySs.
 In the standard model of \citet{b60}, the estimate of the
birthrate of SySs in the Galaxy is 0.076 yr$^{-1}$ and their
number is  close to 3,300, which is similar to our results in the
standard model. Certain differences between the models stem from
the fact that Eq.~(\ref{eq:qcrit}) takes into account the
stabilizing effect of a large mass of stellar core upon RLOF for
stars with convective envelopes that was not considered in YLTK.

The main difference with \citet{b60} is the estimate of the
birthrate of SySs with He WD accretors. In \citet{b60} it is about
0.024 yr$^{-1}$; for CO WD accretors it is about 0.049yr$^{-1}$.
Because of difference in lifetimes, the numbers of systems with
different accretors are more similar: 1560 and 1740. In the
standard model of the present paper, the number of SySs with He WD
accretors is negligible. There are two main reasons for this: 1.
For the same value of $\alpha_{\rm ce}$, the common envelope
formalism of \citet{w84}-- Eq.~(\ref{eq:dek0}), gives larger
reduction of component separation than the YLTK model, leaving no
room for formation of SySs with He WD accretors. In case 4
($\gamma$=1.5), the common envelope model gives a smaller
reduction of binary separation than Webbink's model, then SySs
with He WD accretors appear. 2. In both models, binaries with He
WDs form through common envelopes or RLOF from FGB stars. However,
in  YLTK the upper mass range of the progenitors of He WDs extends
to 2.5 $M_\odot$, while in the present study it is taken as 2.0
$M_\odot$. This difference also allows additional He-accretor
systems to form in simulations of YLTK.

Both studies agree that symbiotic stars are not likely precursors
of SN Ia.

\section{Conclusion}

We performed a detailed study of the formation of symbiotic systems
with WDs as hot components, employing the population synthesis
approach to the evolution of binaries. Several important conclusions
can be drawn:

1. The number of nuclear symbiotic stars in the Galaxy may range
from  about 1,200 to 15,000. These numbers are compatible with
observational estimates. The model birthrate of SySs in the Galaxy
is from
 0.035 to 0.131 yr$^{-1}$.

2. The estimated occurrence rate of symbiotic Novae is between
 1.3 yr$^{-1}$ and 13.5 yr$^{-1}$; weak SyNe
contribute to this rate from about 0.5 to 6.0 events per yr. This
estimate greatly depends on the critical ignition mass of the
hydrogen shell $\Delta M^{{\rm crit}}_{{\rm WD}}$ and accretion
efficiency. It may be compared to the estimate of the Galactic
occurrence rate of classical Novae $41\pm20$\, yr$^{-1}$
\citep{hbf97}. It looks likely that the models with high rates of
symbiotic novae strongly overestimate this rate. This can mean that
(i) the relative fraction of massive accretors is overestimated
because of too steep initial-final mass relation; (ii) the values of
hydrogen-ignition mass $\Delta M^{{\rm crit}}_{{\rm WD}}$ implied by
us  may be underestimated; (iii) mass loss by stellar winds from
accreting dwarfs is underestimated or (iv) accretion efficiency is
overestimated. The latter issue has to be resolved by gas-dynamical
calculations of mass-flows in symbiotic systems. Note also, that
tripling of $\Delta M^{{\rm crit}}_{{\rm WD}}$ in case 11 brings
occurence rate of SyNe to about 1 per yr; having in mind the paucity
of observed SyNe this low number may be preferred. Ignition masses
given by Eq.~(\ref{eq:yunm}) are typically higher than the masses
computed by \citet{b44,b57}. This may mean that  the latter studies
also strongly underestimate $\Delta M^{{\rm crit}}_{{\rm WD}}$.

There  may exist two
varieties of symbiotic stars
--- those with short outbursts and those with long ones.

3. The evolution of symbiotic stars is unlikely to lead to SNe Ia
via accumulation of Chandrasekhar-mass or to the edge-lit
detonations. Helium novae may occur in symbiotic systems if the
energy is efficiently dissipated at the base of accumulated
He-layer.

4. The results of the modeling depend on the assumed
common-envelopes formalism. Within $\alpha$-formalism, variations of
combined parameter $\alpha_{\rm ce}\lambda$ do not strongly affect
model population, since it is dominated by wide systems and systems
that experience stable RLOF. Increase of $\alpha_{\rm ce}\lambda$
shifts the systems produced via channel I into the ranges of
parameters of descendants of channels II and III.

The main effect of applying $\gamma$-formalism is the dominance of
systems formed through common envelopes. Comparison with
observations, whenever possible (Table 2, Figs. 4 -- 11), suggests
that using $\gamma$-formalism it is easier to explain variety of
parameters of observed SySs. This especially concerns existence of
short-period systems and the tendency of SySs for harbouring
relatively low-mass accretors. There is apparent excess of Novae
in $\gamma$-formalism models. However, this may be also an effect
of the underestimate of critical ignition masses or the
overestimate of the number of systems with massive accretors.

5. The wind velocity law affects the occurrence rate of SyNe and the
number of SySs within a factor of 2. Observational data is still not
sufficient to provide real restrictions on the models.

6. To summarize, the models for the population of SySs mostly depend on the
assumed formalism for the common envelope evolution
and critical mass for hydrogen ignition $\Delta M_{\rm
{WD}}^{{\rm crit}}$. These factors have the strongest effect upon
the occurrence rate of SyNe and the total number of SySs,
introducing an uncertainty up to a factor of about 4.

\section*{Acknowledgments}
We are grateful to the referee, C. Tout, for careful reading of
the paper and constructive criticism. We thank Lifang Li, Xuefei
Chen and Xiangcun Meng for some helpful suggestions, and Fenghui
Zhang for providing the data on bolometric corrections.  We
acknowledge J. Miko{\l}ajewska for providing us a compilation of
orbital periods of symbiotic stars. LGL thanks Dr. Richard Pokorny
for correcting English language of the manuscript. This work was
supported by Chinese National Science Foundation under Grants Nos.
10433030 and 10521001, the Foundation of the Chinese Academy of
Sciences (KJCX2-SW-T06) and Russian Academy of Sciences Basic
Research Program ``Origin and Evolution of Stars and Galaxies''.
LRY acknowledges warm hospitality of the Astronomical Institute
``Anton Pannekoek'' where a part of this study was done and
support from NWO and NOVA.

\bsp

\label{lastpage}

\end{document}